\RequirePackage[2020-02-02]{latexrelease}
 \documentclass[twocolumn]{aastex631}
\usepackage{color}
\usepackage{graphicx}
\usepackage{epstopdf}
\usepackage{natbib}
\usepackage{longtable}
\usepackage{amsmath}
\usepackage{rotating}
\usepackage{enumitem}

\newcommand\Wprime{W$^{\prime}$}
\newcommand\rms{\ensuremath{\mathrm{rms}}}
\newcommand\sigmad{\ensuremath{\sigma_\mathrm{MAD}}}
\newcommand\dmean{\ensuremath{d_\mathrm{mean}}}
\newcommand\Ttype{\ensuremath{T_\mathrm{type}}}

\newcommand{\lsim}{\ \raise -2.truept\hbox{\rlap{\hbox{$\sim$}}\raise 5.truept\hbox{$<$}\ }} 
\newcommand{\gsim}{\ \raise -2.truept\hbox{\rlap{\hbox{$\sim$}}\raise 5.truept\hbox{$>$}\ }}

\newcommand\mMbarref{\ensuremath{(\overline{m}-\overline{M})_\mathrm{ref}}}
\newcommand\dref{\ensuremath{d_\mathrm{ref}}}
\newcommand\colref{\ensuremath{col_\mathrm{ref}}}
\newcommand\ustar{\ensuremath{u^*}}
\newcommand\gi{\ensuremath{(g{-}i)}}
\newcommand\gz{\ensuremath{(g{-}z)}}
\newcommand\ui{\ensuremath{(u^{*}{-}i)}}

\newcommand\uz{\ensuremath{(u^{*}{-}z)}}

\newcommand{\feh}{\ensuremath{\mathrm{[Fe/H]}}}
\newcommand\kms{km~s$^{-1}$}

\newcommand\vi{{\ifmmode{(V{-}I)}\else$(V{-}I)$\fi}}
\newcommand\gI{{\ifmmode{(g{-}I)}\else$(g{-}I)$\fi}}
\newcommand\gIacs{{\ifmmode{(g_{475}{-}I_{814})}\else$(g_{475}{-}I_{814})$\fi}}
\newcommand\zbar{\ensuremath{\overline{z}_{850}}}

\newcommand\Mbar{\ensuremath{\overline{M}}}
\newcommand\Mbari{\ensuremath{\overline{M}_i}}

\newcommand\mbar{\ensuremath{\overline{m}}}
\newcommand\mbari{\ensuremath{\overline{m}_i}}

\newcommand\Mibar{\ensuremath{\overline{M}_i}}

\newcommand\MIbar{\ensuremath{\overline{M}_I}}
\newcommand\mibar{\ensuremath{\overline{m}_i}}

\newcommand\kstart{\ensuremath{{k_\mathrm{start}}}}
\newcommand\kend{\ensuremath{{k_\mathrm{end}}}}

\newcommand\lta{\lesssim}
\newcommand\gta{\gtrsim}
\shorttitle{NGVS SBF Distance Catalog}
\shortauthors{Cantiello, Blakeslee, et al.}

\begin{document}

\title{The Next Generation Virgo Cluster Survey (NGVS). III.
~A~Catalog of Surface Brightness Fluctuation Distances and the Three-Dimensional Distribution of Galaxies in the Virgo Cluster}


\author[0000-0003-2072-384X]{Michele Cantiello}
\affiliation{INAF-Astronomical Observatory of Abruzzo, Via Maggini snc, 64020, Teramo, Italy}

\author[0000-0002-5213-3548]{John P. Blakeslee}
\affiliation{NSF’s NOIRLab, 950 N. Cherry Ave., Tucson, AZ 85719, USA}

\author[0000-0002-8224-1128]{Laura Ferrarese}
\affiliation{Herzberg Astronomy and Astrophysics Research Centre, National Research Council of Canada, 5071 W. Saanich Road, Victoria, BC, V9E 2E7, Canada}

\author[0000-0003-1184-8114]{Patrick C\^ot\'e}
\affiliation{Herzberg Astronomy and Astrophysics Research Centre, National Research Council of Canada, 5071 W. Saanich Road, Victoria, BC, V9E 2E7, Canada}

\author[0000-0002-5577-7023]{Gabriella Raimondo}
\affiliation{INAF-Astronomical Observatory of Abruzzo, Via Maggini snc, 64020, Teramo, Italy}

\author[0000-0002-3263-8645]{Jean-Charles Cuillandre}
\affil{AIM Paris Saclay, CNRS/INSU, CEA/Irfu, Universit{\'e} Paris Diderot, Orme des Merisiers, France}

\author[0000-0001-9427-3373]{Patrick R. Durrell}
\affil{Department of Physics and Astronomy, Youngstown State University, Youngstown, OH 44555, USA}

\author[0000-0001-8221-8406]{Stephen Gwyn}
\affil{National Research Council of Canada, Herzberg Astronomy and Astrophysics Research Centre, Victoria, BC, Canada}

\author[0000-0002-3870-1537]{Nandini Hazra}
\affil{Gran Sasso Science Institute: L'Aquila, AQ, IT}
\affil{INAF-Astronomical Observatory of Abruzzo, Via Maggini snc, 64020, Teramo, Italy}

\author[0000-0002-2073-2781]{Eric W.~Peng}
\affiliation{NSF’s NOIRLab, 950 N. Cherry Ave., Tucson, AZ 85719, USA}

\author[0000-0002-0363-4266]{Joel C.~Roediger}
\affiliation{Herzberg Astronomy and Astrophysics Research Centre, National Research Council of Canada, 5071 W. Saanich Road, Victoria, BC, V9E 2E7, Canada}

\author[0000-0003-4945-0056]{R{\'u}ben S{\'a}nchez-Janssen}
\affiliation{UK Astronomy Technology Centre, Royal Observatory, Blackford Hill, Edinburgh, EH9 3HJ, UK}

\author[0000-0001-5208-4697]{Max Kurzner}
\affiliation{Department of Physics and Astronomy, University of Victoria, Victoria, BC V8W 3P2, Canada}

\begin{abstract}
The surface brightness fluctuation (SBF) method is a robust and efficient way of measuring distances to galaxies containing evolved stellar populations. Although many recent applications of the method have used space-based imaging, SBF remains a powerful technique for ground-based telescopes. Deep, wide-field imaging surveys with subarsecond seeing enable SBF measurements for numerous nearby galaxies. 
Using a preliminary calibration, \citet{cantiello18ngvs} presented SBF distances for 89 bright, mainly early-type galaxies observed in the Next Generation Virgo Cluster Survey (NGVS). Here, we present a refined calibration and SBF distances for 278 galaxies extending several magnitudes fainter than in previous work. The derived distances have uncertainties of 5-12\% depending on the properties of the individual galaxies, and our sample is more than three times larger than any previous SBF study of this region. Virgo has a famously complex structure with numerous subclusters, clouds and groups; we associate individual galaxies with the various substructures and map their three-dimensional spatial distribution. Curiously, subcluster A, centered around M87, appears to have two peaks in distance: the main peak at $\sim$16.5~Mpc and a smaller one at $\sim$19.4 Mpc. Subclusters B and C have distances of $\sim$15.8~Mpc. The W and \Wprime\ groups form a filament-like structure, extending more than 15~Mpc behind the cluster with a commensurate velocity increase of $\sim$1000 \kms\ along its length. These measurements are a valuable resource for future studies of the relationship between galaxy properties and local environment within a dynamic and evolving region.

\end{abstract}

\keywords{distance scale --- galaxies: clusters: individual (Virgo) --- galaxies: distances and redshifts}


\section{Introduction} \label{sec:intro}

%

The nearby Virgo cluster of galaxies, along with the groups and filaments that
surround and feed into it \citep[e.g.,][]{tully82} offers a close-up view of the ongoing
assembly of a massive cluster and the transformation of galaxies as they fall into the mix.
The properties of galaxies are influenced by both the large-scale environment
\citep[e.g., the familiar morphology-density and color-density
relations;][]{dressler1980,blanton2005,bamford2009,wang18} and the proximity of neighbors
on smaller scales \citep[e.g.,][]{park2009,ellison2009,ellison2010}.
Virgo is actively accreting both isolated galaxies and small groups. The accreted
galaxies may undergo dramatic changes in morphology and star-formation activity as they
encounter the cluster environment and their host groups are subsequently disrupted
\citep[][]{paudel13,boselli14,lisker2018,benavides2020}. 

Many past studies have divided the galaxies in and around the Virgo cluster into distinct
substructures. The first detailed investigation was probably that of \citet{deVauc61}, who
defined the E, S, W, and X clouds, along with the tighter S$^\prime$, Wa, Wb, and \Wprime\
groups. Cloud E referred to the main body of the Virgo early-type galaxies, while S was the more
dispersed distribution of spirals, and X referred to Virgo's Southern Extension, which was
further split into two components; additional sub-groupings were added by
\citet{deVauc73}. These studies relied on sky positions, velocities, morphologies, and some very crude
distance estimates based on properties such as angular diameters and apparent magnitudes.

A revised nomenclature, describing the same general structures but in finer detail, was
introduced by \citet{binggeli87} based on the Virgo Cluster
Catalog \citep[VCC;][]{binggeli85}. These authors used a combination of velocity and
morphological characteristics to exclude obvious
background galaxies,  then grouped the remaining galaxies based on sky position into
six substructures: cluster~A, centered on the cD M\,87; cluster~B, centered on M\,49, the brightest cluster galaxy (BCG);
cluster~C, centered on the giant elliptical M\,60\footnote{The authors actually reported the center to be M60's smaller neighbor M59.}; the M~cloud, identified as a distinct velocity component by
\citet{ftaclas84}; the W cloud; and the \Wprime\ group \citep[with the latter two  adopted
  from][]{deVauc61}. The VCC did not include the Southern Extension.

In this paper, we refer to the A, B, and C groupings as 
\textit{subclusters}. \citet{binggeli87} described subcluster~C as
being within the elongated halo of A, with a continuous bridge of early-type 
galaxies linking the two concentrations. These substructures were further refined by \citet{binggeli93}, using more extensive radial velocity data. From a combination of the luminosity functions,
radial velocities, and some 21\,cm line-width distance estimates for spirals, these authors
argued that the M, W, and \Wprime\ substructures were significantly more distant than Virgo's
three main subclusters.

To probe the three-dimensional structure of Virgo in more detail, \citet{gavazzi1999} used both fundamental plane (FP) and Tully-Fisher (TF)
distances for over 130 early-type and spiral galaxies. Surprisingly, they found that
subcluster B, associated with M\,49, was about 9~Mpc (one magnitude) more distant than
subcluster~A,  associated with M\,87 (they did not distinguish between A and C). Since the two structures
have nearly the same mean velocity, the interpretation was that B is falling towards A
from behind, at about 760~\kms. Interestingly, the distance to M\,49 itself was much closer
to the mean of subcluster~A than to subcluster B, while the {\it reverse} was true for M\,87. However,
the authors estimated their individual distance errors to be 0.35 and 0.45~mag, or
17\% and 21\%, for the TF and FP distances, respectively, making it difficult to
assign individual galaxies to specific substructures. They further concluded that the M and W
clouds were at roughly {\it twice} the distance of subcluster~A. Several other subgroups that they
identified (i.e., clouds N, S and E) were located close to the mean distance of A, but with
significantly different mean velocities; this could suggest that they are the bound
remnants of groups being accreted by the main cluster.  This work supported the
general view of Virgo as a cluster still in the process of assembly.

A number of studies have explored the relationship between galaxy properties and the local
environment within the extended Virgo cluster region. For instance, \citet{boselli06}
analyzed a sample of 868 UV-selected galaxies in the Virgo region having velocities below
3500~\kms\ to study the role of environment in the
cessation of star formation and the transformation from late- to early-type morphology
 within high-density regions. In addition to  UV and optical photometry, the authors
 made use of extensive mid-infrared and H\textsc{i} radio data. They then studied the distribution of red-sequence, blue star-forming, and transitional ``green valley'' galaxies as a function of the estimated local galaxy density. In particular, they
examined the UV-optical color-magnitude diagrams of subclusters A, B, and C, as well as the
M, W, and \Wprime\ groups, the low-velocity cloud (LVC), and the surrounding field.
As expected, red sequence galaxies predominate within the A and C regions, but also within
the W and \Wprime\ groups. Subcluster~B shows more of a mix of red and blue galaxies, while
the M group, LVC, and surrounding field are dominated by blue galaxies.  More generally, the red fraction correlates with estimated local galaxy density. However, since reliable
distance information was unavailable for most of their sample, the authors made the
assignments of galaxies to substructures based on sky position and velocity, and they used
only projected densities in their analysis, rather than volume densities. Thus, projection effects were a significant limitation in their analysis.

Going further in eliminating contamination to disentangle the various environmental
factors in such a complex region would require precise information on the relative
distances of the galaxies. Precise and reliable distances make it possible to estimate the
local volume density of galaxies, the depth of a galaxy within the cluster well, and the
likelihood of its association with any of the substructures that have been previously identified in
wide-field surveys.  The virial radius of the Virgo cluster (combining the A, B, C
subclusters) is approximately 1.6~Mpc \citep{ferrarese12}, or about one-tenth of the
distance \citep{tonry01,cantiello18ngvs} to the cluster itself. This means that resolving
the internal structure of Virgo would require a distance precision of $\sim\,$5\%.  More compact systems at similar distances (e.g., Fornax), or larger but more distant clusters, would require
correspondingly higher degrees of precision. For example, any attempt to resolve the line-of-sight structure of the Coma cluster at $\sim\,$100~Mpc would be wholly impractical.

But even for Virgo, measuring distances of sufficient precision has been a major
challenge. Cepheids are the premier extragalactic distant indicator, but Cepheid distances are
observationally expensive and cannot be used for the early-type galaxies that dominate the
densest parts of the cluster and its associated groups. The tip
of the red giant branch (TRGB) method is similarly expensive and requires Hubble or JWST image
quality; for this reason, relatively few TRGB distances are available in Virgo (see \citet{bird10}, \citet{leejang17}, \citet{mihos22} and the Appendix of \citet{blakeslee21}).  Type~Ia supernovae are too rare for mapping individual structures in detail, but the planetary nebula luminosity
function (PNLF) method applied using wide-area integral field spectroscopy \citep{spriggs21}
presents an intriguing new possibility;  the globular cluster luminosity function has also been
used to measure distances within Virgo \citep[e.g.,][]{whitmore95b}. Thus far, however, the most detailed work on the
three-dimensional structure of Virgo has been carried out using the SBF method; for a recent review of this technique, see \citet{cantielloblakesleeH0}.


The first attempt to resolve the depth of the Virgo cluster with SBF distances was that of
\citet{tal90} soon after the method was introduced \citep{ts88}. These authors 
showed definitively that NGC~4365 was more distant than the other Virgo giant ellipticals and suggested
that it was a member of the W cloud \citep[now it is recognized as belonging to
\Wprime;][]{mei07xiii,cantiello18ngvs}. However, the stellar population dependence of
the absolute $I$-band SBF magnitude \MIbar\ \citep{tonry91,tonry97} was not recognized
at the time, and the other depth effects reported by \citet{tal90} are attributable to
bluer galaxies having brighter \MIbar\ values. 

A decade later, \citet{west00} used data from the ground-based SBF Survey of Galaxy
Distances \citep{tonry97,tonry01}, fully corrected for stellar population effects, to
constrain the shape and orientation of the main body of Virgo which runs along the axis joining subclusters A and C. 
This study concluded that the fourteen brightest ellipticals in
this central region follow a roughly co-linear distribution in three dimensions along an axis
connecting to a filament that extends in the direction of Abell 1367. \citet{neilsen20}
observed a similar trend using SBF distances measured from Hubble Wide Field Planetary
Camera~2 observations of a dozen bright Virgo galaxies along the A-C subcluster axis; they also
confirmed that NGC\,4365 (located south of this axis) is significantly more distant.  
\citet{west00} did not consider
subcluster~B or the W/\Wprime\ groups in their analysis, but they showed that NGC\,4168,
 which is now known to be a member of the M~group \citep{cantiello18ngvs}, lay along an extension of the same ``principal axis". However, the study was limited by the relatively small number of sample galaxies
and the $\sim\,$10\% errors for these ground-based distance measurements.

\citet{mei07xiii} examined the detailed structure of the Virgo cluster using SBF distances
for 84 galaxies observed in the ACS Virgo Cluster Survey \citep[ACSVCS;][]{cote04}. Thus,
their sample size was much larger, and the mean random distance errors of the space-based
measurements were more than a factor of two smaller, than for the data set analyzed by
\citet{west00}. Five of the galaxies (including NGC\,4365) were found to be members of the
\Wprime\ group, about 6~Mpc beyond the main cluster. Excluding these objects, the {\it rms} depth of the
remaining 79 galaxies, including members of the A, B, and C subclusters, was estimated to
be just $\sigma_d=0.6\pm0.1$ Mpc. Analysis of the three-dimensional positions of these
galaxies showed that they defined a triaxial distribution with
axis ratios 1:0.7:0.5 inclined about 30$^\circ$ from the line of sight. SBF distances
measured from ACSVCS data for several additional Virgo galaxies, and a slightly revised
SBF calibration, were presented by \citet{blake09}. Despite the enormous advance over
earlier studies of Virgo's structure, the targeted ACSVCS covered only a small fraction of
the cluster area, missing many galaxies suitable for SBF distance measurements;
for instance, the ACSVCS sample did not include any galaxies in the M or W clouds. Ideally, one would
like high-quality imaging over the entire cluster.

The Next Generation Virgo Cluster Survey \citep[NGVS][]{ferrarese12} is a deep, multi-band
imaging survey with the MegaCam instrument on the Canada–France–Hawaii Telescope (CFHT) covering
an area of 104 square degrees and extending out to the virial radii of the A and B
subclusters.  The NGVS has already produced numerous results on the properties, varieties,
and distributions of Virgo galaxies and their globular cluster systems
\citep[e.g.,][]{durrell14,ferrarese16,powalka16,sanchez16,sanchez19,roediger17,liu20,lim20},
as well as other objects along the line of sight
\citep[e.g.,][]{chen13,raichoor14,Lokhorst16,fantin17}. The best seeing conditions during
the acquisition of NGVS imaging were reserved for the $i$ band, and this makes
it ideal for SBF measurements.

As a first step to mapping the three-dimensional structure of Virgo from NGVS,
\citet{cantiello18ngvs} published SBF measurements for 89 VCC galaxies with total $B$
magnitude $B_T\lesssim 13$ mag and provided a
preliminary calibrations for \Mibar\ as a function of various colors and color combinations
that could be constructed from the NGVS $\ustar,g,i,z$ passbands. The broadest baseline
\uz\ color provided the best overall calibration with the lowest scatter. In a few cases
where \ustar\ photometry was not available, a calibration that combined both \gi\ and
\gz\ was used. The sample included five galaxies in the M and W clouds, both of
which lie at roughly 30~Mpc but are widely separated on the sky, and a similar
number in the \Wprime\ group at $\sim\,$22.5~Mpc.

That study demonstrated the potential of the method in mapping the
structure of the Virgo cluster and its surrounding groups using NGVS
data. However, a more extensive sample encompassing fainter galaxies
is desirable due to the significantly larger number of available
targets, enabling a more detailed examination of the cluster's
internal structure. Expanding to fainter galaxies necessitates
refining the calibration, particularly towards the blue colors typical
of low-mass galaxies. A significant advantage of NGVS lies in its
ability to employ a single dataset comprehensively and
consistently, covering a range of masses (and colors) from bright galaxies reaching $M_B\sim-23$ mag, 
like M\,87 or M\,60, to much fainter ones with $M_B\sim-12$ mag.

This paper presents SBF measurements and calibrated distances for about 280
galaxies observed as part of NGVS. This is the largest and most comprehensive sample of
SBF measurements in Virgo to date, extending to fainter and significantly more diverse
galaxies than those analyzed by \citet{cantiello18ngvs}.  The following section, \S\ref{sec_sample}, summarizes the sample of galaxies that we study, while \S\ref{sec_measure} describes the SBF
measurements and \S\ref{sec_analysis} presents the refined SBF calibration based on this
larger sample and the resulting distances.  We find 
a median relative distance uncertainty of 7.5\% for bright galaxies with $B_T<13$ mag,
and ${\sim\,}10\%$ for fainter galaxies ($B_T>13$~mag).
The final section, \S\ref{sec_philosophy}, discusses the three-dimensional distribution of the 
sample galaxies and
the effect of internal stellar population gradients on the SBF magnitudes, with a
comparison to theoretical models. We conclude with some thoughts on the prospects for going beyond
the current work.

\section{The Galaxy Sample} \label{sec:style}
\label{sec_sample}


As with the first SBF distance paper in the NGVS series
\citep[][NGVS Paper XVIII, hereafter referred to as Paper~I]{cantiello18ngvs},
the present study is based on CFHT/MegaCam imaging data from the NGVS.  
Full details on the NGVS survey observations and image processing are presented in
\citet{ferrarese12}. As a brief summary, the NGVS exploits the capabilities of CFHT/MegaCam to reach 5-$\sigma$ limiting magnitudes for point sources of 26.3, 26.6, 25.8, and 24.8~mag in the
stacked \ustar, $g$, $i$, and $z$ images, respectively, across the 104~deg$^2$ area of the survey.
These limits are well beyond the turnover magnitude for the globular cluster (GC) luminosity function \citep[e.g.,][]{durrell14},
even when the GCs are superimposed on a bright galaxy background. Moreover, the median FWHM for each NGVS band ($u^*giz$) is better than 0\farcs9, with a median FWHM of 0\farcs54 in the $i$ band. These properties make the NGVS images extremely well suited to an SBF analysis, as most of the potentially
contaminating sources can be identified and removed.

For this second NGVS paper on SBF distances, 
we have analysed $\sim300$ galaxies brighter than $B_T\approx 19$ mag (median $B_T=14.5$ mag) in the 
Virgo Cluster Catalogue (VCC) of \citet{binggeli85}. This includes approximately half a dozen faint galaxies ($B_T\sim18$
mag) that were not included in the VCC, but are available in the NGVS catalog of \citet{ferrarese20}. Although we imposed no strict selection on morphology, we did choose galaxies that exhibit some degree of morphological regularity. Such regularity is an essential requirement for conducting a meaningful analysis of fluctuation amplitudes \citep[see, e.g.,][]{cantielloblakesleeH0}. A future paper in the NGVS series will introduce a customized morphological classification system for NGVS galaxies, including the $\sim$ 300 galaxies that make up our current sample. For the time being, we simply note that the sample is dominated by galaxies with ``early-type" morphologies. For example, 184 of the 215 galaxies having quality classification codes of $q1$ or $q2$ (see the following section) are classified as E or ES galaxies (usually corresponding to the VCC classifications of E, S0, dE or dS0). The remaining systems have later type morphologies but usually with the presence of a prominent bulge or bar.
We point out that this requirement for some degree of morphological regularity could introduce environment-related selection effects in our sample, since galaxies within dense groups and clusters tend to have less star formation and a smoother appearance.

Table \ref{tab_data} lists the main properties of the galaxies included in the present work.
For each galaxy, we list the VCC number from \citet[][]{binggeli85} (Col. 1)  its alternate and IAU name (Cols. 2, 3);
J2000 celestial coordinates in degrees (Cols. 4, 5); total $B_T$ magnitude from the VCC (Col. 6);
galactic extinction \citep[][Col. 7]{sfd98}; 
heliocentric velocity $v_h$ from the NASA Extragalactic Database (Col. 8); morphological \Ttype\ (Col. 9) and its error (given in  parenthesis) from Hyperleda \citep{makarov14}. 

The SBF analysis for galaxies in the present sample is carried out as detailed in Paper~I and briefly described in the following sections.


\begin{deluxetable*}{lllcccccc}
\tabletypesize{\scriptsize}
\tablecaption{List of targets with SBF measurements (Excerpt)}
\tablehead{
  \colhead{VCC} &
  \colhead{Alt. Name} &
  \colhead{IAU Name} &
  \colhead{RA (J2000)} &
  \colhead{Dec (J2000)} &
  \colhead{$B_T$} &
  \colhead{$E(B{-}V)$} &
  \colhead{$v_{h}$} &
  \colhead{$T$} 
  \\
  \colhead{   } &
  \colhead{         } &
  \colhead{         } &
  \colhead{   (deg)  } &
  \colhead{  (deg)    } &
  \colhead{(mag)} &
  \colhead{ (mag)           } &
  \colhead{\kms       } &
  \colhead{          } 
  } \startdata
      (1)&   (2)      &     (3)      &        (4)   &    (5)   &  (6)   &    (7)  &      (8)  & (9)   \\   
\hline
      32 & IC0767    & NGVSJ12:11:02.73+12:06:14.4 & 182.7613724 &   12.1039918 &    14.00 &  0.027 &    1877 &   -5.0 (  1.0) \\
      33 & IC3032    & NGVSJ12:11:07.76+14:16:29.3 & 182.7823394 &   14.2748006 &    14.70 &  0.038 &    1179 &   -4.0 (  1.8) \\
      49 & NGC4168   & NGVSJ12:12:17.26+13:12:18.7 & 183.0719332 &   13.2051960 &    12.20 &  0.037 &    2273 &   -4.9 (  0.4) \\
      66 & NGC4178   & NGVSJ12:12:46.57+10:51:59.8 & 183.1940241 &   10.8666242 &    11.90 &  0.028 &     374 &    6.9 (  1.2) \\
      89 & NGC4189   & NGVSJ12:13:47.24+13:25:29.5 & 183.4468247 &   13.4248521 &    12.50 &  0.033 &    2098 &    5.9 (  0.9) \\
      92 & NGC4192   & NGVSJ12:13:48.29+14:54:01.8 & 183.4512262 &   14.9005060 &    10.90 &  0.035 &    -142 &    2.6 (  0.8) \\
     107 & PGC039071 & NGVSJ12:14:09.75+13:14:00.5 & 183.5406147 &   13.2334632 &    19.40 &  0.030 & \nodata &   -5.0 (  2.0) \\
     140 & IC3065    & NGVSJ12:15:12.56+14:25:58.4 & 183.8023507 &   14.4328799 &    14.30 &  0.036 &     995 &   -2.2 (  1.2) \\
     145 & NGC4206   & NGVSJ12:15:16.82+13:01:26.4 & 183.8200770 &   13.0239966 &    12.80 &  0.032 &     703 &    4.0 (  0.3) \\
     157 & NGC4212   & NGVSJ12:15:39.34+13:54:05.4 & 183.9138976 &   13.9015055 &    11.90 &  0.033 &     -88 &    4.9 (  0.8) \\
     167 & NGC4216   & NGVSJ12:15:54.39+13:08:58.0 & 183.9766077 &   13.1494566 &    11.00 &  0.032 &     131 &    3.0 (  0.5) \\
     199 & NGC4224   & NGVSJ12:16:33.79+07:27:43.4 & 184.1407813 &    7.4620674 &    12.90 &  0.024 &    2584 &    1.0 (  0.3) \\
     200 & PGC039331 & NGVSJ12:16:33.71+13:01:53.7 & 184.1404477 &   13.0315970 &    14.70 &  0.030 &       8 &   -4.3 (  1.6) \\
     220 & NGC4233   & NGVSJ12:17:07.68+07:37:27.4 & 184.2820056 &    7.6242759 &    13.00 &  0.024 &    2275 &   -2.0 (  0.5) \\
     222 & NGC4235   & NGVSJ12:17:09.88+07:11:29.7 & 184.2911844 &    7.1915747 &    12.70 &  0.019 &    2263 &    1.0 (  0.4) \\
     226 & NGC4237   & NGVSJ12:17:11.43+15:19:26.5 & 184.2976068 &   15.3240384 &    12.50 &  0.030 &     864 &    4.0 (  0.5) \\
     230 & IC3101    & NGVSJ12:17:19.65+11:56:36.5 & 184.3318761 &   11.9434644 &    15.20 &  0.028 &    1430 &   -5.0 (  1.4) \\
     \multicolumn{8}{c}{\nodata}
\\
\enddata
\tablecomments{This table is available in its entirety in a machine-readable form in the online journal. A portion is shown here for guidance regarding its form and content.}
\end{deluxetable*}
\label{tab_data}

\begin{deluxetable*}{llcccccccccccccccc}
\tabletypesize{\scriptsize}
\tablecaption{SBF, colors and distances for the 278 galaxies in our sample (Excerpt)}
\tablehead{
  \colhead{VCC} &
  \colhead{\uz} &
  \colhead{\gi} &
  \colhead{\gz} &
  \colhead{$\overline{m}_i$} &
  \colhead{\mMbarref} &
  \colhead{\dref} &
  \colhead{\dmean} &
  \colhead{Area} &
  \colhead{$\langle Rad\rangle$} &
  \colhead{Flag} 
  \\
  \colhead{   } &
  \colhead{(mag) } &
  \colhead{(mag)} &
  \colhead{(mag)} &
  \colhead{ (mag)          } &
  \colhead{      (mag)                      } &
  \colhead{ (Mpc)    } &
  \colhead{ (Mpc)    } &
  \colhead{(arcmin$^2$)} &
  \colhead{ (arcsec)    } &
   \colhead{} 
 } \startdata
      (1)&   (2)      &     (3)      &        (4)   &    (5)   &  (6)   &    (7)  &      (8)     &      (9)           &            (10)   &    (11)           \\   
\hline
      32 &    2.152 ( 0.031) &    0.801 ( 0.008) &    1.009 ( 0.031) &   29.788 ( 0.057) &   31.133 ( 0.091) &    16.85 ( 0.71) &    16.95 ( 0.90)  &  0.85 & 23.8  &	q2   \\
      33 &    2.140 ( 0.038) &    0.819 ( 0.010) &    0.991 ( 0.038) &   29.489 ( 0.143) &   30.855 ( 0.162) &    14.82 ( 1.11) &    14.85 ( 1.22)  &  1.34 & 27.4  &	q1   \\
      49 &    2.623 ( 0.007) &    0.983 ( 0.002) &    1.214 ( 0.007) &   31.628 ( 0.102) &   32.539 ( 0.128) &    32.20 ( 1.90) &    32.92 ( 1.83)  &  4.11 & 51.5  &	q2   \\
      66 &    1.032 ( 0.270) &    0.374 ( 0.046) &    0.125 ( 0.269) &   28.583 ( 0.162) &   31.405 ( 0.814) &    19.10 ( 7.17) &    19.40 ( 6.23)  &  0.73 & 60.7  &	q3   \\
      89 &    2.011 ( 0.069) &    0.768 ( 0.020) &    0.978 ( 0.069) &   31.004 ( 0.363) &   32.475 ( 0.376) &    31.26 ( 5.42) &    31.37 ( 5.51)  &  0.36 & 63.6  &	q3   \\
      92 &    \nodata        &    0.908 ( 0.017) &    1.190 ( 0.061) &   29.931 ( 0.163) &   31.056 ( 0.203) &    16.26 ( 1.52) &    16.26 ( 1.52)  &  0.78 & 78.6  &	q3   \\
     107 &    1.768 ( 0.147) &    0.707 ( 0.036) &    0.753 ( 0.145) &   29.158 ( 0.375) &   30.848 ( 0.424) &    14.78 ( 2.89) &    14.49 ( 2.82)  &  0.03 &  3.8  &	q3   \\
     140 &    2.270 ( 0.014) &    0.924 ( 0.004) &    1.120 ( 0.014) &   29.934 ( 0.093) &   31.188 ( 0.114) &    17.28 ( 0.90) &    16.95 ( 1.00)  &  1.31 & 27.2  &	q1   \\
     145 &    1.430 ( 0.535) &    0.791 ( 0.203) &    0.729 ( 0.533) &   28.557 ( 1.563) &   30.657 ( 1.855) &    13.53 (11.57) &    11.87 (10.14)  &  0.39 & 42.7  &	q3   \\
     157 &    1.802 ( 0.071) &    0.764 ( 0.017) &    0.966 ( 0.071) &   30.066 ( 0.136) &   31.727 ( 0.175) &    22.15 ( 1.78) &    21.33 ( 1.82)  &  0.85 & 67.1  &	q3   \\
     167 &    2.501 ( 0.030) &    0.999 ( 0.013) &    1.234 ( 0.029) &   30.303 ( 0.227) &   31.336 ( 0.242) &    18.50 ( 2.06) &    18.18 ( 2.01)  &  1.05 & 59.2  &	q2   \\
     199 &    2.568 ( 0.012) &    0.989 ( 0.005) &    1.211 ( 0.010) &   30.761 ( 0.082) &   31.710 ( 0.113) &    21.97 ( 1.15) &    21.94 ( 1.08)  &  0.38 & 31.2  &	q2   \\
     200 &    2.328 ( 0.033) &    0.854 ( 0.014) &    1.073 ( 0.032) &   29.815 ( 0.169) &   31.011 ( 0.183) &    15.93 ( 1.35) &    16.14 ( 1.47)  &  1.36 & 27.5  &	q1   \\
     220 &    2.737 ( 0.008) &    1.033 ( 0.004) &    1.302 ( 0.007) &   31.751 ( 0.145) &   32.491 ( 0.164) &    31.49 ( 2.39) &    31.81 ( 2.34)  &  1.02 & 32.9  &	q2   \\
     222 &    2.545 ( 0.015) &    0.965 ( 0.006) &    1.209 ( 0.012) &   31.233 ( 0.122) &   32.196 ( 0.145) &    27.49 ( 1.84) &    27.60 ( 1.78)  &  0.37 & 28.0  &	q2   \\
     226 &    2.171 ( 0.042) &    0.816 ( 0.012) &    1.069 ( 0.041) &   30.040 ( 0.107) &   31.372 ( 0.131) &    18.81 ( 1.14) &    19.06 ( 1.31)  &  0.64 & 54.5  &	q2   \\
     230 &    2.267 ( 0.014) &    0.875 ( 0.007) &    1.064 ( 0.014) &   30.167 ( 0.158) &   31.415 ( 0.171) &    19.18 ( 1.51) &    19.17 ( 1.62)  &  0.27 & 12.7  &	q1   \\
    \multicolumn{11}{c}{\nodata}
\\
\enddata
\tablecomments{This table is available in its entirety in a machine-readable form in the online journal. A portion is shown here for guidance regarding its form and content. All magnitudes are  AB~mag.}
\label{tab_measures}
\end{deluxetable*}

\section{SBF measurements}
\label{sec_measure}

For measuring the SBF amplitudes in the sample galaxies, we adopted the same procedures developed and described in Paper~I, which are based on well established methods
\citep{bva01,blake09,blake10b,cantiello05,cantiello07b,cantiello07a,cantiello11a,cantiello13,mei05iv,mei05v,moresco22,cantielloblakesleeH0}. 
A few minor changes were made, as detailed below.

\begin{figure*}
\begin{center}
\includegraphics[scale=.28]{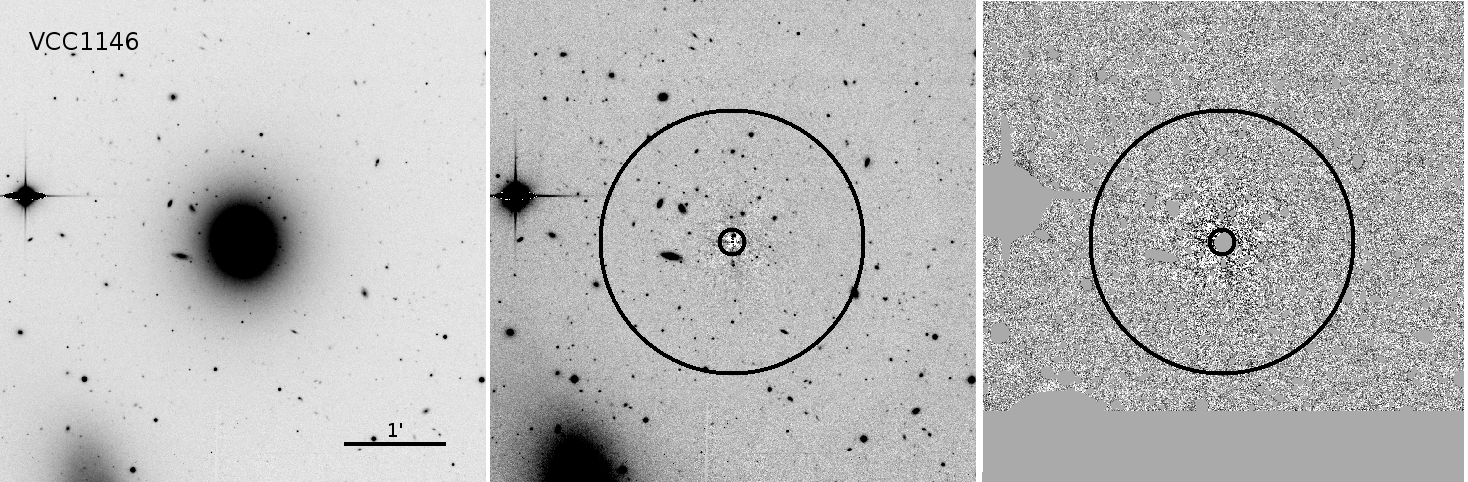} \\
\vspace{0.5cm}
\hspace{-0.5cm}
\includegraphics[scale=.75]{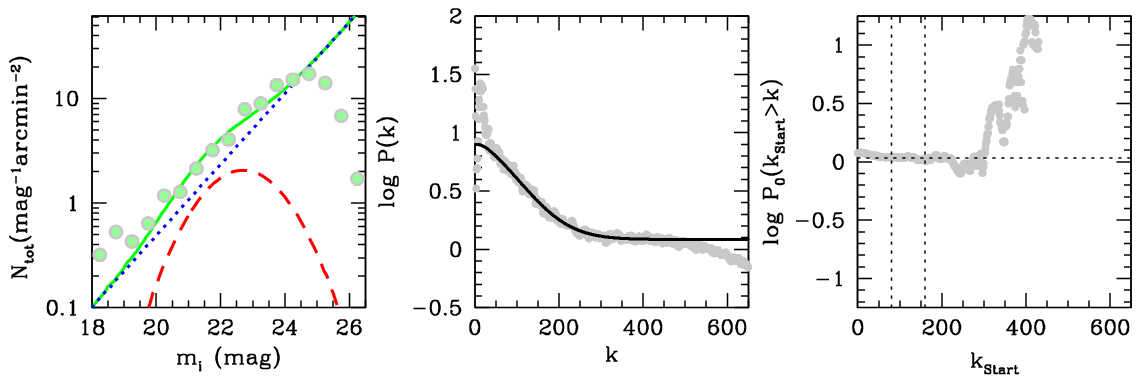}\\
\end{center}
\caption{SBF analysis images and plots for one of our sample galaxies (VCC\,1146).
    Starting from upper left: $i$-band image, residual, and
   residual masked image (first to third panel). The black annuli in
   the second and third panels show the inner and outer radii of the
   region adopted for SBF measurements. Lower left panel: Fitted
   luminosity function of external sources. Filled green circles mark
   observational data, with a downturn at faint magnitudes due to incompleteness; the solid green 
   line represents the best fit to the data,  after accounting for incompleteness;
    the two components of the total luminosity function, i.e. the
   background galaxies and the GC luminosity function, are shown with blue dotted and
   red dashed curves, respectively. Lower central panel: azimuthal average of
   the residual image power spectrum (gray dots) and the fit obtained
   according to the procedure described in text (solid black curve).
Lower right panel: the fitted $P_0$ value as a function of the lowest wavenumber \kstart\ used for the fit (the highest wavenumber used in all the fits is chosen as $\kend{\,=\,}450$). The vertical lines at 80 and 160 show a factor-of-two range in \kstart\ over which there is  little variation in the fit, and we adopt the median $P_0$ (indicated by the dotted horizontal line) over this range; see text for details.
\label{pofit}}
\end{figure*}

In broad terms, the SBF distance measurement entails the following steps: 
\begin{enumerate}[label={\it \alph*.}]
\item First, we subtract the local sky background, mask bright stars 
and neighboring galaxies, then model the 2-D galaxy surface brightness distribution 
as a series of elliptical isophotes and subtract this model from the image.
Large-scale model residuals are then fitted with a bicupic spline and subtracted
to produce a clean residual image. In terms of the SBF definition,
which involves the ratio of the second to the first moment of the
luminosity function of the observed stellar population \citep{ts88}, this
process generates a model image representing the first moment of the
light distribution and a residual image related to the second moment
of the light distribution.

\item We then detect all point-like and extended sources
(foreground stars, GCs in the target galaxy, faint background
galaxies) down to a fixed signal-to-noise threshold using an automated photometry
program. These sources, along with other contributors to non-SBF
variance (e.g., visible dust, brighter satellite galaxies, tidal
features, regions of poor model residuals), are then combined in a
global mask. This masking stage, along with the previous step of
galaxy modeling, subtraction, and a low-order fit to the background,
follows an iterative procedure.

\item Next we obtain one or more accurate point spread function (PSF)
templates for the image.  This is a crucial step because stellar
fluctuations are convolved with the PSF. In the Fourier domain, these
fluctuations multiply the Fourier transform of the PSF (convolved with the window function of the mask;
see item $d$ below). Thus, accurate measurement of the SBF amplitude requires
a robust determination of the PSF. To ensure the required reliability,
we generate several PSF templates from stars in the field, normalized to unit flux.
Generally, for NGVS, we find that empirical PSF templates provide
more robust results (i.e., a better match to the azimuthally averaged power-spectrum)
compared to model PSFs.

\item We then analyze the power spectrum of the masked
residual frame, normalized to the square root of the
galaxy model. After azimuthally averaging the power spectrum,
$P(k)$, the total fluctuation amplitude corresponds to the $P_0$ coefficient
in the fitted equation:
$P(k)=P_0 \times E(k)+P_1\,,$
where $E(k)$ is the ``expectation power spectrum,'' calculated by convolving the power spectrum of
the normalized PSF with the power spectrum of the mask window function, 
and the $P_1$ term is independent of wavenumber $k$
in the absence of correlated noise.

\item The fitted $P_0$ includes contributions from all astrophysical sources
of fluctuation, both the stellar SBF that we seek to measure and 
contamination from faint sources below the detection limit. 
The presence of dust could potentially impact
the power spectrum and, consequently, the fitted $P_0$. However,
the $u^*$ and $g$ NGVS images make it possible identify the dusty regions in color maps,
so that the SBF measurements can be confined to clean areas.

Because of their high frequency and radial concentration in early-type galaxies, 
GCs below the direct detection threshold
are the main source of contamination in the $P_0$ measurement.
The surface density of background galaxies,
being lower and relatively uniform, has a less dramatic impact and is
more easily constrained. The ``residual power'' $P_r$ is estimated from
sources fainter than the detection limit (which varies with
galactocentric radius) by extrapolating the fitted combined GC and
background galaxy luminosity function, multiplied by the square of
the source flux, from the radially-dependent detection limit down to
zero flux (see Figure \ref{pofit},  lower left panel). The residual power $P_r$ 
is subtracted from $P_0$ to obtain intrinsic stellar fluctuations,
typically denoted as $P_f=P_0{-}P_r$ from which the SBF magnitude is derived
as $\mbar={-}2.5\log P_f + m_\mathrm{zp}$, where the zero-point term is
$m_\mathrm{zp}=30$ mag in the NGVS images.

\item Adopting a value for $\Mbar$ from either an empirical or
theoretical SBF calibration \citep[generally based on galaxy color;
see][for a discussion]{blake12b}, we finally obtain the distance modulus 
$(\overline{m}{\,-\,}\overline{M})$.
\end{enumerate}

Figure~\ref{pofit} illustrates the basic steps in the SBF measurement.
The top panels show the image of the target galaxy (VCC\,1146 in this example), the 
residual frame (step~$a$ in the list above) after galaxy subtraction, and the residual frame after masking the regions contaminated by foreground/background sources and image defects (step~$b$).
The lower panels show the fitted background luminosity function model (left panel; step~$e$), 
the power spectrum of the residual frame compared to the scaled PSF power spectrum
(middle panel; step $d$), and the fitted $P_0$ value as a function of the starting 
wavenumber \kstart\ for the fit to the power spectrum.

The upturn in the power spectrum at low wavenumber $k$ (lower middle panel) is due to the imperfect subtraction of large-scale features in the residual frames (step $a$). The downturn at high $k$ occurs because of the correlation of noise in adjacent pixels caused by the sinc-like interpolation kernel during image stacking \citep[see][]{tal90,cantiello05,mei05iv}. For these reasons, the lowest ($k<\kstart$) and highest ($k>\kend$) wavenumbers are excluded from the power spectrum fits. The downturn at high $k$ from the noise correlation is straightforward to determine, and here we conservatively stop each fit at a maximum wavenumber of $\kend{\,=\,}450$. However, the fitted $P_0$ is more sensitive to excess power at low
$k$, and this depends to some degree on the quality of the isophotal fit. As an example,
each dot in the lower-right panel of Figure~\ref{pofit} shows the (logarithm of) the $P_0$ value obtained when starting the fit at \kstart\ in the range $0<\kstart<450$.
When using $\kstart{\,\lesssim\,}80$, the fitted $P_0$ values show an upward trend with decreasing 
\kstart, a consequence of the excess power due to imperfect galaxy model residuals on large scales.

Compared to Paper~I, we have made a small change in the procedure for determining $P_0$. In the previous work, we examined the $\log P_0$ versus \kstart\ plots (lower right panel in Figure~\ref{pofit}) to visually identify the best values of \kstart\ for stable $P_0$ determinations, and averaged over
a representative range. 
For the present measurements, we follow the approach described by \citet{blake97}. First 
we identify the value of \kstart\ where the $\log P_0$ versus \kstart\ relation stops declining; 
call this $k_{s,1}$. We then perform a series of fits with \kstart\ in the range 
$k_{s,1}\leq\kstart\leq k_{s,2}$, where $k_{s,2}{\,\simeq\,}2\,k_{s,1}$.
For example, in Figure~\ref{pofit}, we use $(k_{s,1},k_{s,2}) = (80,160)$, 
typical values for the galaxies we analyze.
We then take the median of the fitted $P_0$ values from among all these fits, and 
estimate the error by combining the $P_0$ scatter in quadrature with median fit uncertainty.
This approach, as compared to basing the range purely on visual inspection, results in 
a more homogeneous set of $P_0$ values for the sample galaxies.
However, in practice, it negligibly changes the \mibar\ measurements
presented in Paper~I, with a median difference in the final \mibar\ values
$\Delta({\mibar}_\mathrm{,new}-{\mibar}_{\mathrm{,Paper~I}})<0.01$~mag. 

Another change with respect to Paper~I is that we now present a quality classification code, or flag, for each SBF measurement: $q1$ for excellent quality; $q2$ for good quality; $q3$ for low quality. These flags are assigned based on our confidence in the result of the full SBF analysis, considering the severity of the isophotal model residuals, possible effects of bright nearby objects, the degree of masking, and the regularity of the power spectrum (which can be affected by the masking, chip defects, isophotal irregularity, etc.).

We also measure the integrated colors of the galaxies in the same annuli
used for the SBF measurements. Table \ref{tab_measures} collects the results of the SBF analysis, and the distances derived as detailed in the forthcoming section \S~\ref{sec_analysis}. In particular, we report the
\uz, \gi\ and \gz\ colors and their uncertainties (Cols.~2-4)\footnote{Colors and SBF magnitudes in the Table are uncorrected for Galactic extinction.}; SBF magnitude (Col.~5);  
  the preferred distance modulus and the distance in Mpc (Cols. 6 and 7); the mean distance using all four SBF versus color calibrations derived here (Col. 8); the area and the median radius of the annulus adopted for each galaxy in our sample (Cols. 9 and 10); and the quality flag for the SBF measurement (Col.~11).

\begin{figure*}
\includegraphics[width=0.95\textwidth]{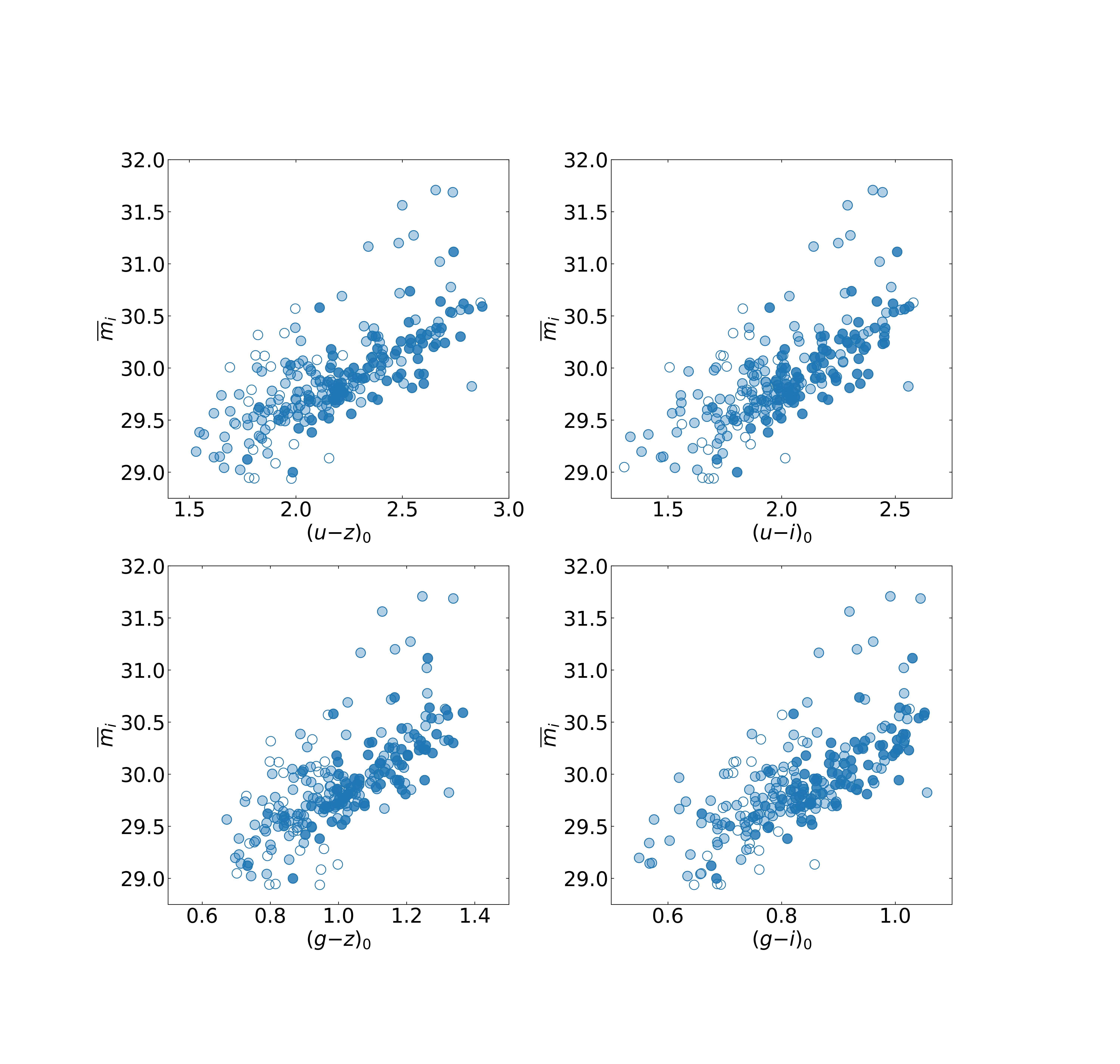} 
\vspace{-1.5cm}
\caption{Measured \mibar~magnitudes versus color, for the full sample of 278 galaxies reported in Table \ref{tab_measures}. Different symbols show galaxies flagged with different quality classification codes: dark-filled symbols ($q1$), light-filled symbols ($q2$) and 
open blue circles ($q3$). The median error bars are $\Delta \mibar=0.09$, 0.11
and 0.22 mag for $q1$, $q2$ and $q3$ sources, respectively.
\label{mbarcol}}
\end{figure*}

In all the figures that follow, the  colors and magnitudes reported in
   Table \ref{tab_measures} have been extinction corrected using the values from \citet{sfd98}\footnote{The $E(B{-}V)$ in Table \ref{tab_data} were converted into extinction in each band using the following coefficients: $A_{\lambda}/E(B{-}V)$=4.594/3.560/2.464/1.813/1.221 for $u^*$/$g$/$r$/$i$/$z$, determined with the York Extinction Solver using the \citep{fitzpatrick99} extinction law.}. 
Figure \ref{mbarcol} plots the SBF magnitudes for the sample of
galaxies reported in Table \ref{tab_measures} versus various colors. In the panels of the figure, different shades of blue refer to different quality 
classification codes: i.e., dark-filled symbols for $q1$, light-filled symbols for $q2$ and open circles for $q3$. In all, we measured SBF distances for 278 galaxies\footnote{The original sample had 295 galaxies but 17 were rejected for poor quality due to a variety of reasons: e.g., the faintness of the target, the presence of very bright contaminants, poor fitting results, etc.}.

\section{Analysis}
\label{sec_analysis}
To derive distances from SBF measurements, an accurate calibration of
$\overline{M}_i$ is required (step $f$ in the previous section). In this section, we describe the
calibration procedure adopted for the new sample of galaxies, discuss the 
differences with respect to the calibrations given in Paper~I (and other relations available in the literature) and compare the empirical calibrations to those predicted by stellar population models.

\subsection{Calibrating absolute SBF magnitudes}
\label{sec_cal}
As shown in Figure \ref{mbarcol}, the relation between \mibar\ and the galaxy integrated colors (with \mibar\ becoming fainter as the color gets redder) appears even with no correction for the presence of bright galaxies in the background of Virgo, including members of the more distant substructures of the cluster, like the W or \Wprime~clouds (see section \S~\ref{sec_distances}). To model the SBF versus color relation, we adopted a procedure similar to the approach of \citet[][ACSFCS-V hereafter]{blake09} for the analysis of SBF measurements from HST/ACS data of Fornax and Virgo cluster galaxies. The procedure takes into account the cluster depth, as well as the mean uncertainties on the measured SBF, and rejects the outliers with a $\sigma$-clipping iteration. 

In more detail, using the data in Table~\ref{tab_measures} and plotted in Figure~\ref{mbarcol}, we first selected the targets flagged as $q1$ or $q2$ and with uncertainty $\Delta \mibar \leq0.25$ mag, to avoid targets with less reliable measurements. The selected sample contained $N_{sel}=212$ galaxies. We then fitted the \mibar\ versus color relations with a clipping algorithm to reject the outliers (mostly galaxies in the W and \Wprime~clouds). For the fitting, we used the $astropy$ \citep{astropy22} class $FittingWithOutlierRemoval$, which performs a fit according to the chosen model and clips the outliers at each iteration, until no outliers remain. We chose to use a polynomial model function and tested degrees between one and five; for the clipping, we used \sigmad, a {robust estimate of the scatter} derived from the median absolute deviation (MAD; $\sigmad= 1.48\,$MAD), with a $3\sigma$ rejection parameter. 

We adopted a third-order polynomial because, on average for the four colors, it gave a slightly better $\chi^2_{\nu}\sim1.0$ and lower scatter compared to lower and higher degree polynomials. This is consistent with previous studies. For example, in ACSFCS-V, a third degree polynomial proved the most adequate description of the \zbar\ versus \gz\ relation.

\begin{figure*}
\includegraphics[width=0.95\textwidth]{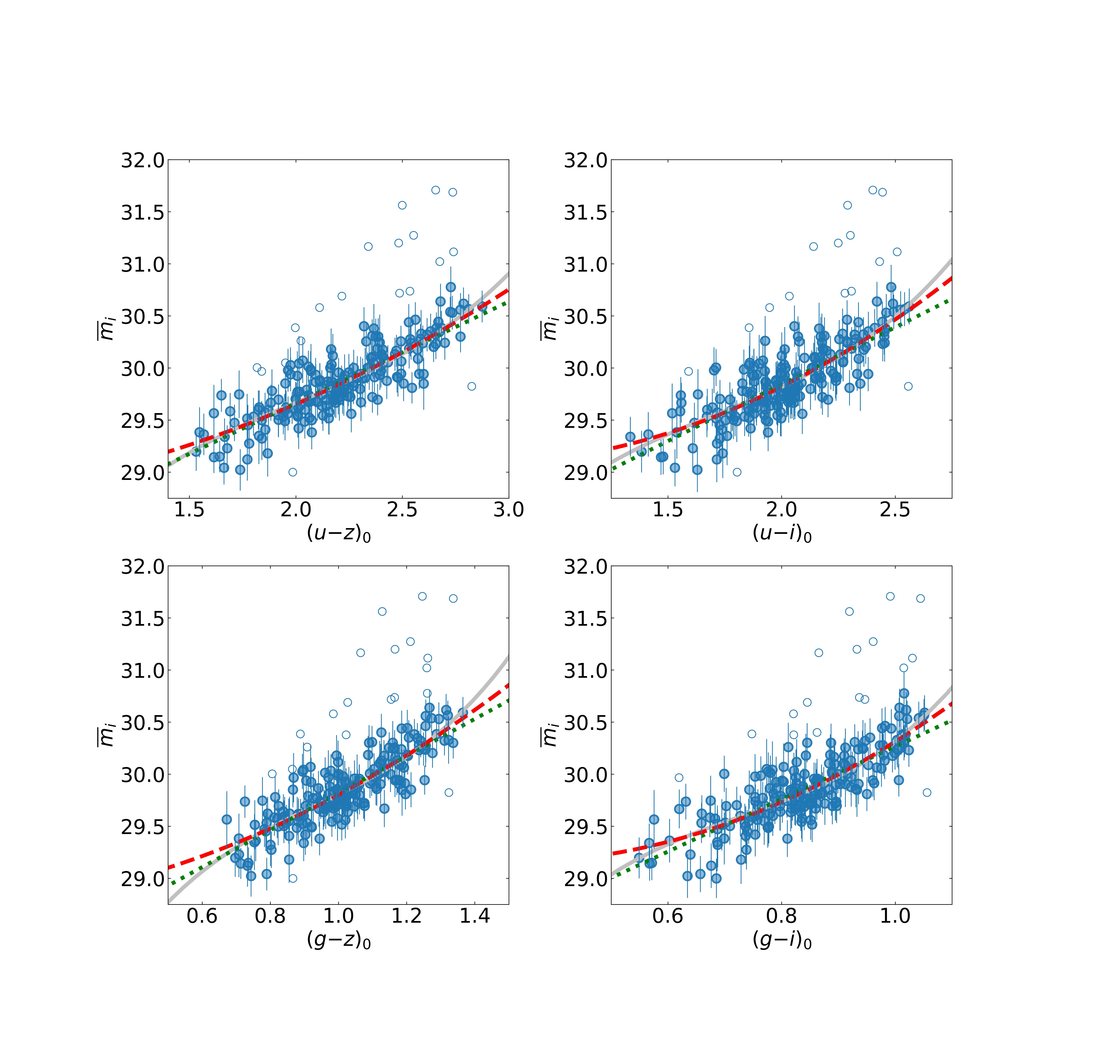} 
\vspace{-1.5cm}
\caption{Illustration of the fitting and $3\sigma$ clipping procedures for the different colors. The 212 targets with code $q1$ and $q2$ measurements are used. The open circles show the sources rejected after the clipping iteration. The polynomials fitted to data are shown as dotted green (linear fit), dashed red  (second degree) and solid gray (third degree) curves. 
\label{fitpoly}}
\end{figure*}

After just three iterations, the clipping and fitting procedure converged in all four colors considered. The numbers of non-rejected galaxies used in the fitting procedure were $N_{fit}=193,196,192,195$ for the \uz, \ui, \gz, and \gi\ relations, respectively.  The reduced $\chi_\nu^2$ in all cases was $\sim1.0$, with $\rms$ scatter of $\sigma_0\sim$0.18 mag for the \uz~ and \gz~ relations, and $\sigma_0\sim$0.20 mag for the other two colors. 
The results of the polynomial clipping and fitting procedure are shown in Figure~\ref{fitpoly}. 

We adopt a two-step procedure to determine the absolute zero points of the \Mbari\ 
versus color relations. Similar to the approaches used by
ACSFCS-V and \citet[][]{mei07xiii}, we start by assuming the Cepheid-calibrated
mean SBF distance modulus of Virgo
from \citet{tonry01}, modified as described in \citet{blake02} and then reduced by 0.023 mag for
consistency with the revised LMC distance of \citet{pietrzynski19} 
giving a fiducial distance modulus of 31.067~mag, with a systematic uncertainty of 0.09~mag
(see \citet{blakeslee21} for further discussion).
We subtract this value from the zero points of the apparent \mbari\ versus color relations
to get a first estimate of the zero-point calibration.

Next, to ensure the best consistency of our distances with previously published 
high-quality SBF distance measurements, 
we refine the zero points of the fitted SBF-color relations by comparing the first-step
calibrated distances with  measurements for the same individual galaxies tabulated
in ACSFCS-V (again adjusted for the revised LMC distance modulus).
In total, we find 83 galaxies in common between our sample and with ACSFCS-V for the \uz\ and \ui\ 
calibrations, and 85 for \gz\ and \gi.
The small difference in sample size is due to the fact that some NGVS targets lack $u^*$-band data.
Using these sets of galaxies in common between NGVS and ACSFCS-V, we do final shifts 
in the zero points of the calibration relations so that the median difference between 
the ACS and NGVS SBF distance moduli is equal to zero for each color calibration.

Table \ref{tab_fit} provides the results of the cubic polynomial calibration equations 
for each color in the form:
\begin{equation}
    \Mbari=a_0 x^3 +a_1 x^2+ a_2 x+ a_3, 
\end{equation}
where $x\equiv col-\colref$, $col$ is the color index used for the calibration,
and $\colref$ is the reference color, taken as the median color 
(rounded to the nearest 0.05 magnitude bin) in that index for the galaxies in common with ACSFCS-V.
The absolute zero points, $\Mbari(x{=}0) = a_3$, are set by the comparison to ACSFCS-V, as described above.

Thanks to the large color range of the galaxies in our sample, we have characterized the intrinsic scatter for both the blue and red side of the SBF versus color relations, using $\colref$ as blue/red separation limit. In Table \ref{tab_fit} we report some other results from the fitting procedure for each color: the scatter $\sigma_0$ of the fitted curve, the reduced $\chi^2_{\nu}$, the intrinsic {\it cosmic} scatter in the SBF calibration $\sigma_{cos}$, together with $\sigma_{cos,blue}$ and $\sigma_{cos,red}$, the cosmic scatter derived for the galaxies bluer/redder than $\colref$. 

To determine the cosmic scatter\footnote{The cosmic scatter definition we adopt here is the irreducible scatter
in SBF magnitudes at fixed galaxy properties (specifically, the color,
in our case, as we calibrate \Mbari\ against the galaxy color). By
definition, this represents the difference in \Mbari\ one might
measure from ideal observations for the SBF signal in two different
galaxies with the same color, due to the intrinsic difference
in the luminosity function of their stellar populations \citep[see][]{tonry00}}, $\sigma_{cos}$, for each relation,  we first estimated the depth of Virgo, $\sigma_{est}$, adopting the same approximations used in ACSFCS-V (their eq.~3):
\begin{equation}
    \sigma_{est}=\sqrt{\frac{1}{2}(\sigma_{R.A.}^2+\sigma_{Dec.}^2)}\times \frac{\pi}{180} \times \frac{5}{\ln 10}~ \mathrm{mag}, 
\end{equation}
based on the positional scatter ($\sigma_{R.A.}$ and $\sigma_{Dec.}$) of the sample of Virgo cluster galaxies from the VCC. After removing galaxies with $B_T{\,>\,}18$ mag or with $v_{h}>3000$ \kms, we find $\sigma_{est}=0.12$ mag. 

We then estimated the cosmic scatter in the \mibar--color relations as
\begin{equation}
    \sigma_{cos}=\sqrt{\sigma_0^2-\sigma_{est}^2-\sigma_{err}^2},
\end{equation}
where $\sigma_{0}$ is the observed scatter reported in Table \ref{tab_fit}, and $\sigma_{err}$ is the median measurement error on \mibar. We find $\sigma_{cos}\sim0.07,0.06$ mag for \uz, \gz, respectively, and slightly larger values for \ui\ and \gi. Values of $\sigma_{cos}\lsim0.08$ mag are consistent with the upper limits of previous estimates \citep{tonry00,blake09}. Larger scatters may be explained by the wider color and magnitude ranges of our sample compared to previous studies. For example, the ACSVCS-V \gz\ calibration sample includes a total of 14 galaxies in Virgo and Fornax with $\gz<1.1$ mag; in our sample we have $\sim200$ galaxies bluer than this. Using the same fitting procedures above, and further limiting the sample to class $q1$ and bright class $q2$ galaxies ($N_{sel}\sim120$, $N_{fit}\sim110$) to better match with the magnitudes and color ranges of previous SBF calibration studies, we find $\sigma_{cos,red}\lsim0.06$ mag for the \gz\ and \gi\ calibrations, and $\sigma_{cos,red}\sim0.08$ for \uz\ and \ui. The \gz\ result agrees well with that of ACSFCS-V. Such a conservative selection, though, narrows the color range by $\sim20\%$ on the blue side, greatly reducing the usefulness of the \Mbari\ calibrations for dwarf galaxies.

The increased SBF scatter at blue colors is expected due to the more complex age and metallicity variations
in the stellar populations of low-mass, blue galaxies compared to bright, massive ones \citep[e.g.][]{carlsten19,greco21,kim21,moresco22}. Any contribution of very young stars especially increases the scatter of the SBF magnitude at a given color; most often, the scatter is observed to increase in moving from massive red ellipticals to bluer, low-mass dwarfs. Interestingly, in the case of \uz, the scatter is actually smaller on the blue side, suggesting that this broadest baseline color is more effective in characterizing the stellar populations of  blue galaxies.

Combining the \mbari\ values and the colors in Table \ref{tab_measures}, with the equations in Table \ref{tab_fit}, four different estimates of distance can be obtained for each galaxy in our sample. As in Paper~I,  we adopt the \uz\ calibration for reference\footnote{Although the \Mibar-\uz\ relation has slightly larger $\sigma_{cos}$ compared to \gz, the smaller fitting uncertainties, and shallower dependence of \Mibar\ on the color, typically result in a smaller total error on the distance.}, using the mean of the \gi\ and \gz\ distances for the targets with no $u^*$-band data. 
The reference distances and distance moduli are reported in Table \ref{tab_measures}, indicated with the \textit{ref} label. In the table, we also provide the distance \dmean\ obtained by averaging the estimates from all four calibrations (or two, in cases where no $u^*$ photometry is available).

Finally, the uncertainties reported on the distance moduli in Table \ref{tab_measures} are the square sum of the measurement errors $\Delta \mbari$, the propagation of the color uncertainty on the cubic polynomial, and the cosmic scatter $\sigma_{cos}$ (adopting $\sigma_{cos}=\sigma_{cos,red}$ for galaxies with color larger than $\colref$, and $\sigma_{cos}=\sigma_{cos,blue}$ for galaxies with colors $col<\colref$).

While extensive literature exists on the distances of galaxies in our
sample, a fair comparison would necessitate rescaling all available
distances to a common reference calibration. Moreover, given the
wealth of literature on Virgo, the number of available distance estimates
likely surpasses a thousand measurements with diverse and often poorly
characterized uncertainties.
A comprehensive comparison, encompassing numerous galaxies with
diverse distance estimates and standardizing to the same zero point,
is beyond the scope of this work.

Nevertheless, we recognize the importance of comparing our SBF
distances with those available from the TRGB method. The preference for TRGB distances is
due to their reliability and potential for providing a Cepheid-independent
calibration of the SBF method, useful for comparison to the Cepheid-based distance ladder.
Adopting the TRGB distances for M\,87 (VCC\,1316, NGC\,4486) and M\,60 (VCC\,1226, NGC\,4472) 
from \citet{bird10} and \citet{leejang17}, homogenized as in \citet{blakeslee21},
we find a very close agreement between our distance moduli in Table \ref{tab_measures} and literature
TRGB value for M\,87, with $\Delta(SBF-TRGB) = 0.03 \pm 0.13$ mag, and a 1-$\sigma$ match of
$\Delta (SBF-TRGB) = 0.13\pm 0.13$ mag for M\,60.

\begin{deluxetable*}{lcccccccccccc}
  \label{tab_fit}
  \tablecaption{Coefficients and results of the fits $\Mibar=a_0 x^3 +a_1 x^2+ a_2 x+ a_3$, where $x\equiv col-col_\mathrm{ref}$, and $col$ is the photometric
  color used for the particular calibration.} 
\tablewidth{0pt}
 \tablehead{
\colhead{Color} & \colhead{$a0$} & \colhead{$a1$} & \colhead{$a2$} & \colhead{$a3$} & \colhead{$\chi^2_{\nu}$} & \colhead{$\sigma_0$} & \colhead{$\sigma_{cos}$} & \colhead{$\sigma_{cos,blue}$} & \colhead{$\sigma_{cos,red}$} & \colhead{\colref} &\colhead{$N_\mathrm{fit}$}}\startdata
$u{-}z$ &  0.422 (0.364) &  0.418 (0.216) &  0.981 (0.075) & $-$1.135 (0.020) &  1.02 &  0.18 &  0.07 &  0.06 &  0.08 &   2.35  &   193 \\
$u{-}i$ &  0.537 (0.532) &  0.683 (0.326) &  1.165 (0.087) & $-$1.135 (0.022) &  0.99 &  0.19 &  0.10 &  0.11 &  0.08 &   2.15  &   196 \\
$g{-}z$ &  3.092 (2.600) &  1.500 (0.807) &  1.788 (0.143) & $-$1.126 (0.020) &  1.04 &  0.17 &  0.06 &  0.06 &  0.05 &   1.10  &   192 \\
$g{-}i$ & 10.085 (6.832) &  5.204 (2.042) &  2.816 (0.192) & $-$1.129 (0.022) &  1.01 &  0.19 &  0.09 &  0.11 &  0.03 &   0.90  &   195 \\
\enddata
\end{deluxetable*}



\subsection{Comparison with Paper~I and previous calibrations}

Paper~I presented the detailed SBF measurement procedure and derived a calibration for the SBF versus color relation, using a sample of 32 galaxies with accurate distances from ACSFCS-V\footnote{There are 36 galaxies in common between ACSFCS-V and Paper~I, but we excluded those in the \Wprime~cloud or those with no available $u^*$-band data.}. Due to the relatively small color range, it was appropriate in that case to fit the color dependence of \Mbari\ with a linear relation. In addition to the different fitting scheme used in Paper~I, we adopt here a revised LMC distance and a richer sample of galaxies to set the zeropoint: 85 galaxies for \gz~and \gi, and 83 for \uz~and \ui.

\begin{figure*}
\includegraphics[width=0.9\textwidth]{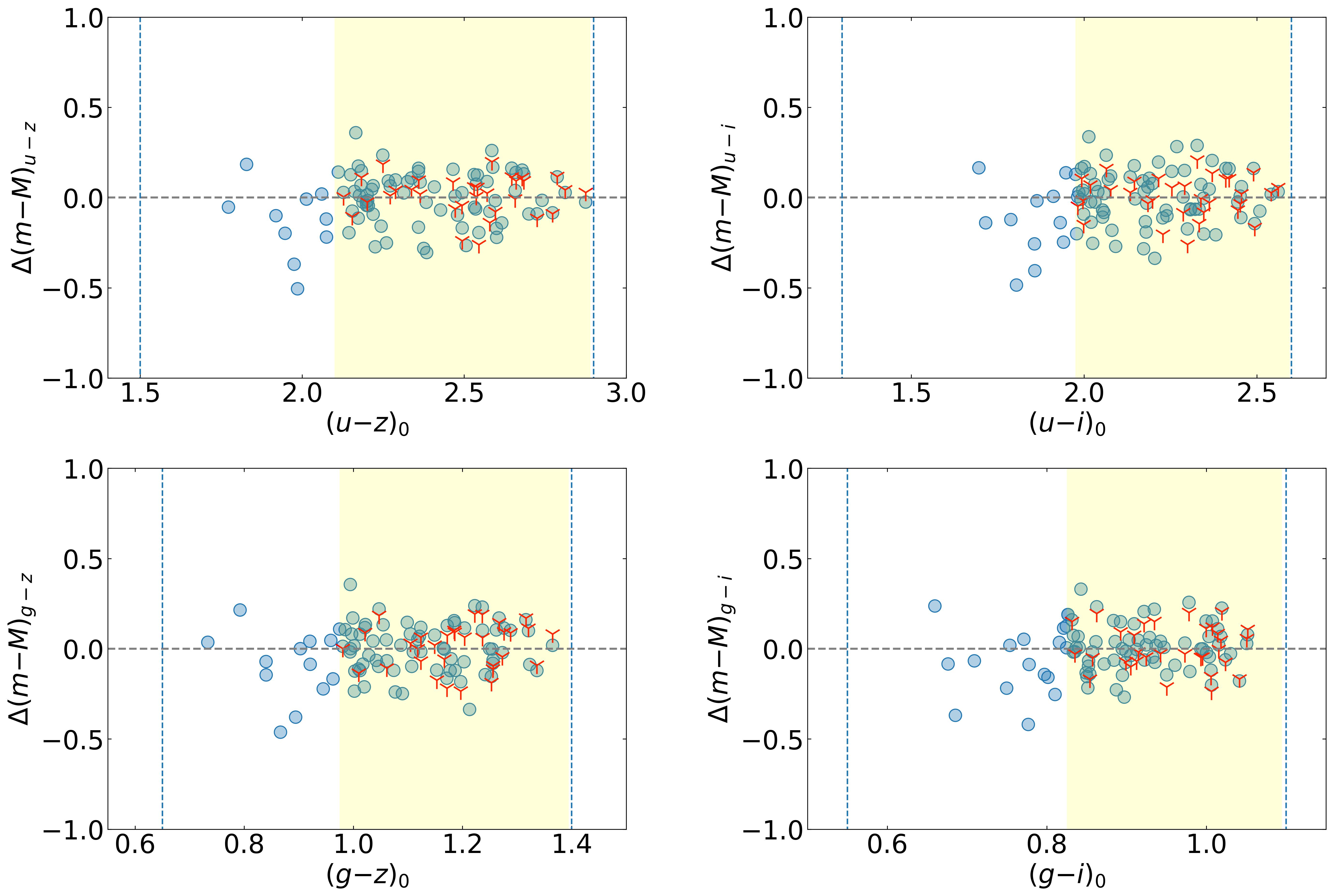} 
\caption{Distance modulus differences between the present NGVS and the ACSFCS-V distances ($\Delta (m{-}M)=(m{-}M)_{NGVS}-(m{-}M)_{ACSFCS-V}$), used to set the zeropoint of the calibration equations. Red three-pointed stars mark the sample used in Paper~I, while the yellow shaded area highlights the color range of the data used in Paper~I. Filled blue dots mark the new calibration sample. The vertical lines refer to the range of validity of the revised calibration equations presented in this work. The color interval for the present sample extends toward bluer color than the sample of calibrators because of the fitting approach adopted (see Section \ref{sec_cal}).
\label{deltacs}}
\end{figure*}

Figure \ref{deltacs} shows the difference in distance moduli for galaxies in common between ACSVCS-V and the present set of NGVS measurements, as well as the earlier Paper~I values, for all four of the color calibrations derived. Note the few calibrators in Paper~I toward the blue edge  of the sample at \uz$\sim2.2$ (the red symbols in the figure) and the lack of blue calibrators at, e.g., \uz$\leq2.1$ mag. As a consequence, the useable color range for the relations presented in Paper~I is considerably smaller, typically about one half of the new updated interval (see Figure~\ref{deltacs}, with the yellow shaded regions for Paper~I and the vertical dashed lines for this work).

\begin{figure*}
\includegraphics[width=0.9\textwidth]{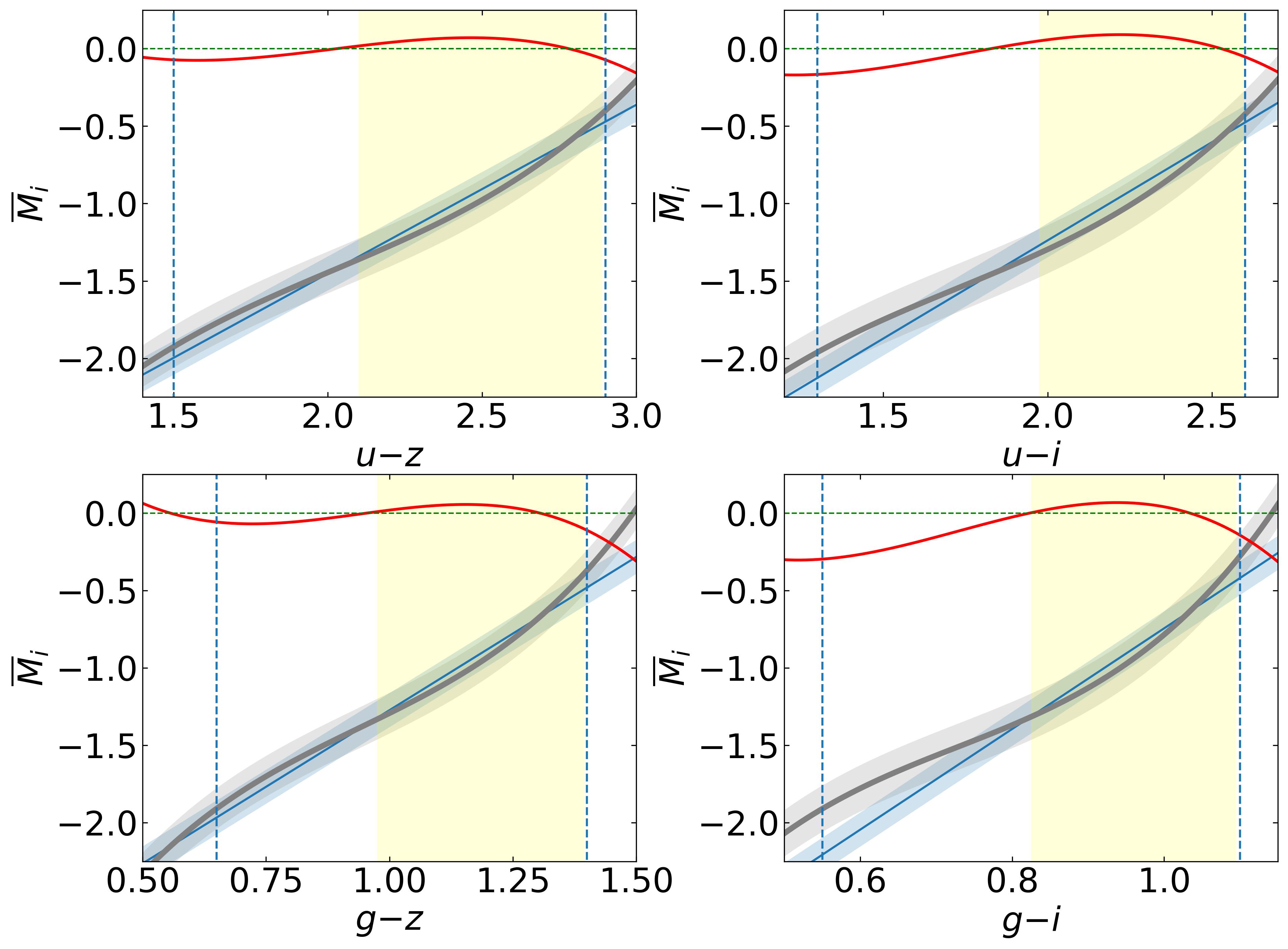} 
\caption{A comparison of the calibration equations derived in this work (gray curves and area) and the ones presented in Paper~I (blue solid lines and shaded area, extrapolated outside the yellow shaded area). The yellow shaded regions and vertical dashed blue lines are as in Figure \ref{deltacs}. The red curve shows the difference between the two calibration equations (Paper~I minus present). The green horizontal line is the zero-difference level, reported for sake of clarity. 
\label{compacal}}
\end{figure*}

As a further comparison, Figure~\ref{compacal} shows the two sets of calibrations (gray for the present relations; blue for Paper~I) along with the difference between the two (red curve in the figure). The largest difference in the predicted \Mibar\ between the relations and, therefore, in the derived distances, is observed over distinct ranges for each color --- for the \gi\ and \gz\ relations toward the very red end of the color range, where the upturn in the third degree polynomial fit diverges sensibly from the linear trend; toward intermediate colors with respect to the range of validity of Paper~I (the yellow shaded regions in the figure, at \uz$\sim2.5$ mag) for \uz~ and \ui, and at the very blue tail of the \ui\ and, most notably, the \gi\ relations.  

The preference of the third degree polynomial over a linear relation at red colors cannot be robustly constrained with the present data, as the difference between the two relations starts to diverge toward colors more red than the range spanned by our sample. As already explained, our preference for the third degree polynomial is based on the $\chi_{\nu}^2$ values and the scatter of the fits. Moreover, even for the very red objects in our sample, over the color range where both relations are valid, the separation between the fitted relations is consistent within our estimated scatter, with a higher divergence observed in the \gi\ color.

However, for the blue side of the relations, especially for the \gi\ calibration, the linear extrapolation from Paper~I (blue solid lines beyond the yellow shaded areas in Fig.~\ref{compacal}) would imply  overly large mean distances for blue galaxies. Hence, a measurable difference between a linear and a third-degree fit emerges. For the specific case of the \gi~relations, we direct the reader to Section \ref{sec_gi}. 

In conclusion, the combination of factors reported in the above paragraphs further 
support our choice for the degree of the fitting polynomial. We emphasize that
the distances presented here supersede those presented in Paper~I.


\subsection{Comparison with other blue calibrations}
\label{sec_gi}
\citet{carlsten19} and \citet{kim21} derived the \Mbari\ versus \gi\ color relation from an analysis of low-mass galaxies. As anticipated, such targets have intrinsically bluer colors compared to more massive galaxies. 
The two independent relations are:
$(i)$ Eq.~(4) from \citet{carlsten19}, 
$\Mibar=(-3.17 \pm 0.19) + (2.15\pm0.35) \times \gi$, 
with $\rms=0.26$ mag, which is valid over the color range $0.3 \leq \gi \leq 0.8$; and 
$(ii)$ Eq.~(2) from  \citet{kim21}, 
$\Mibar=(-2.65 \pm 0.13) + (1.28\pm0.24) \times \gi$, 
with estimated $\rms=0.16$ mag for the color range $0.2 \leq \gi \leq 0.8$. 
These equations are tied to TRGB distances; for homogeneity with our calibration, we shift the zero points of these relations fainter by 0.023~mag to be consistent with the revised LMC distance modulus.

The \citeauthor{carlsten19} calibration refers to the same instrument and passbands used in this work (MegaPrime/MegaCam), hence no further passband transformation is required to compare it with our data. The \citeauthor{kim21} relation is instead derived from Subaru/Hyper Suprime-Cam data, which use a photometric system slightly different from that of CFHT. We use the equations derived by \citeauthor{kim21} from PARSEC isochrones \citep{bressan12} and, like these authors, assume that the $i$-band magnitudes of the two systems are almost identical, while for the $g$-band a correction is required. To this aim, we invert the equation given in section 5 of  \citeauthor{kim21} and derive $\gi_{HSC}=1.045\gi_{CFHT}+0.006$. Hence, the revised \Mibar\ versus \gi\ in the CFHT passbands is:   
$\Mibar=-2.64 + 1.34 \times \gi$.

\begin{figure}
\includegraphics[width=0.45\textwidth]{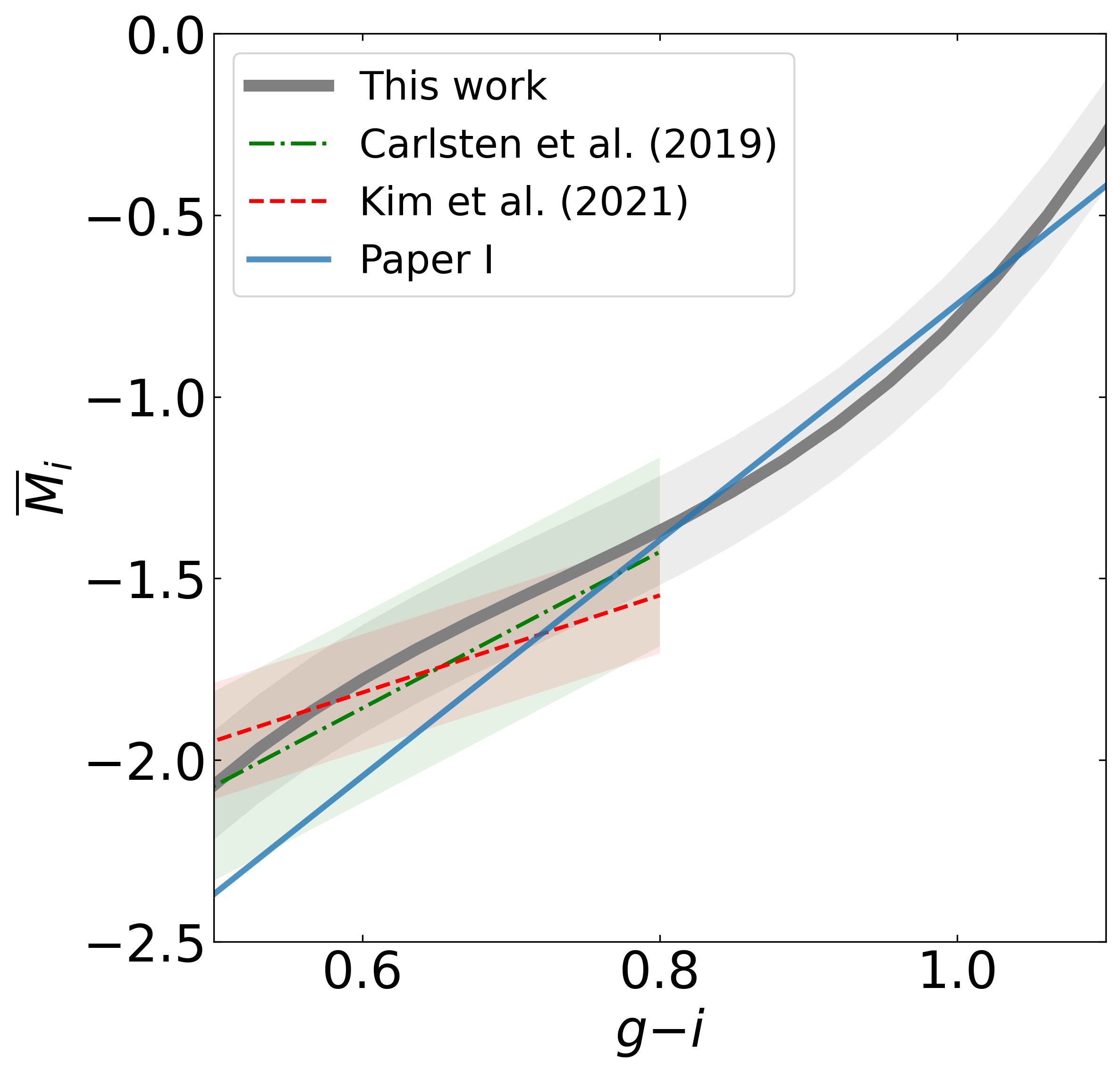} 
\caption{A comparison of absolute SBF calibrations for the relation derived in this work (gray solid curve), and the calibrations from \citet[green dot-dashed curve]{carlsten19} and \citet[red dashed curve]{kim21}. The shaded areas show the $\pm 1\sigma$ regions for each calibration within its range of validity. 
The solid blue line shows the calibration equation from Paper~I extrapolated toward the wider color interval of this paper.
\label{fitblue}}
\end{figure}

The relations are plotted in Figure \ref{fitblue}, along with our best-fit equation. 
In the figure, our \gi\ calibration is illustrated with an error region $\rms=0.15$ mag, obtained from the quadrature sum of $\sigma_{cos}\gi$ and the median measurement error in $\mibar$. 
In spite of the differences between the three equations --- for example, in the approach used for data analysis (e.g., the way the $P_r$ correction is handled), in the standard candle used to anchor the calibration (TRGB for for the literature relations; Cepheids for our relation), or in the typical targets used --- the relations overlap within their $\rms$ scatter over the range of validity in color.

In Figure \ref{fitblue} we also show, with a blue line, the extrapolation of the Paper~I \Mibar\ versus \gi\ relation derived for colors \gi$\gsim0.8$ mag. The plot clearly evidences the systematically too bright SBF amplitudes extrapolated for \gi$\leq0.7$ mag, independent of the calibration with which they are compared. As discussed in the previous section, this evidence supports our preference for a higher degree polynomial over the wide color range inspected in this work, compared to the linear relation adopted in previous studies, like Paper~I.

\subsection{SBF calibrations vs simple stellar population models}

As discussed in the previous section, obtaining a distance from \mbar\ requires an estimate of \Mbar.  This typically involves an empirical calibration that parameterizes the correlation between \Mbar\ and the properties of the unresolved stellar population in the galaxy via its integrated color. However, theoretical predictions of \Mbar\ from stellar population synthesis models also represent a viable alternative for deriving absolute SBF versus color relations. 

There have been many efforts in the literature to predict absolute SBF magnitudes
using stellar population models 
\citep[e.g.,][]{worthey93a,bva01,raimondo05,cook20,chung20,greco21}.
SBF magnitudes are heavily weighted toward the brightest evolutionary stages,
and theoretical calibrations of the SBF method from different sets of models 
can differ by several tenths of a magnitude or more, particularly in near-infrared passbands
where luminous stars in fast evolutionary stages---especially the thermally pulsating 
asymptotic giant branch \citep[e.g.,][]{raimondo09}---are still inadequately
modeled. Nevertheless, despite the disagreement
among model predictions (or, arguably, because of it),
SBF magnitudes, gradients, and distance-independent SBF
``fluctuation colors'' \citep{cantiello05,cantiello07a,jensen15,rb21} 
hold great promise as tools for scrutinizing the properties of 
integrated stellar populations within galaxies, complementary to more traditional 
spectrophotometric indicators.

In this section, we focus on the role of model SBF predictions in illuminating the physical properties
underlying the empirical calibration equations derived in the previous section.
With some exceptions \citep[e.g.,][]{biscardi08}, the vast majority of published SBF distances rely on empirical calibrations \citep{tonry01,bva01,blake09,cantiello18gw,jensen03,jensen21}. In this section, we present a comparison of our empirical SBF–color relations with predictions from the Teramo-SPoT simple stellar population (SSP) models \citep[][]{cantiello03,raimondo05, raimondo09}.

The SPoT SBF models are based on a stellar population synthesis code first described and tested observationally by \citet{brocato99, brocato00}. The code is optimized to reproduce both the observed distribution of stars in the color-magnitude diagrams and the integrated properties (color and spectral indices, energy distribution, etc.) of star clusters and galaxies. A detailed description of the SPoT models and ingredients can be found in literature cited above. The updated color and SBF models used here, and in Paper~I, are reported in Table \ref{tab_models}.

\begin{deluxetable*}{lcccccc}
\tablecaption{Integrated colors and SBF magnitudes from the Teramo-SPoT Simple Stellar Population models.}
\tablehead{
  \colhead{[Fe/H]} &
  \colhead{Age (Gyr)} &
  \colhead{$u{-}i$} &
  \colhead{$g{-}i$} &
  \colhead{$u{-}z$} &
  \colhead{$g{-}z$} &
  \colhead{$\overline{M}_i$} 
  } \startdata
      (1)&   (2)      &     (3)      &        (4)   &    (5)   &  (6)   &    (7)     \\   
\hline
   -0.35 &      1 &     1.518 &     0.588 &     1.676 &     0.746 &    -2.083  \\
   -0.35 &      2 &     1.730 &     0.707 &     1.878 &     0.855 &    -1.581  \\
   -0.35 &      3 &     1.887 &     0.789 &     2.052 &     0.954 &    -1.482  \\
   -0.35 &      4 &     1.961 &     0.825 &     2.135 &     0.999 &    -1.348  \\
   -0.35 &      6 &     2.044 &     0.858 &     2.221 &     1.035 &    -1.344  \\
   -0.35 &      8 &     2.143 &     0.907 &     2.339 &     1.103 &    -1.084  \\
   -0.35 &     10 &     2.214 &     0.937 &     2.418 &     1.141 &    -1.035  \\
   -0.35 &     12 &     2.256 &     0.952 &     2.466 &     1.162 &    -0.996  \\
   -0.35 &     14 &     2.309 &     0.974 &     2.525 &     1.190 &    -0.977  \\
   \hline
    0.00 &      1 &     1.638 &     0.640 &     1.813 &     0.815 &    -1.262  \\
    0.00 &      2 &     1.950 &     0.804 &     2.154 &     1.008 &    -1.067  \\
    0.00 &      3 &     2.113 &     0.880 &     2.339 &     1.106 &    -0.950  \\
    0.00 &      4 &     2.218 &     0.925 &     2.459 &     1.166 &    -0.859  \\
    0.00 &      6 &     2.332 &     0.968 &     2.574 &     1.210 &    -0.796  \\
    0.00 &      8 &     2.427 &     1.009 &     2.686 &     1.268 &    -0.732  \\
    0.00 &     10 &     2.505 &     1.041 &     2.777 &     1.313 &    -0.586  \\
    0.00 &     12 &     2.575 &     1.068 &     2.855 &     1.348 &    -0.497  \\
    0.00 &     14 &     2.638 &     1.094 &     2.925 &     1.381 &    -0.485  \\
    \hline
    0.40 &      1 &     1.889 &     0.762 &     2.104 &     0.977 &    -0.828  \\
    0.40 &      2 &     2.120 &     0.867 &     2.351 &     1.098 &    -0.448  \\
    0.40 &      3 &     2.282 &     0.934 &     2.525 &     1.177 &    -0.263  \\
    0.40 &      4 &     2.377 &     0.973 &     2.629 &     1.225 &    -0.149  \\
    0.40 &      6 &     2.507 &     1.027 &     2.770 &     1.290 &    -0.191  \\
    0.40 &      8 &     2.614 &     1.070 &     2.887 &     1.343 &    -0.115  \\
    0.40 &     10 &     2.705 &     1.109 &     2.992 &     1.396 &    -0.008  \\
    0.40 &     12 &     2.775 &     1.139 &     3.070 &     1.434 &     0.090  \\
    0.40 &     14 &     2.837 &     1.165 &     3.139 &     1.467 &     0.118  \\
\label{tab_models}
\enddata
\end{deluxetable*}

\begin{figure*}
\begin{center}
\includegraphics[width=0.8\textwidth]{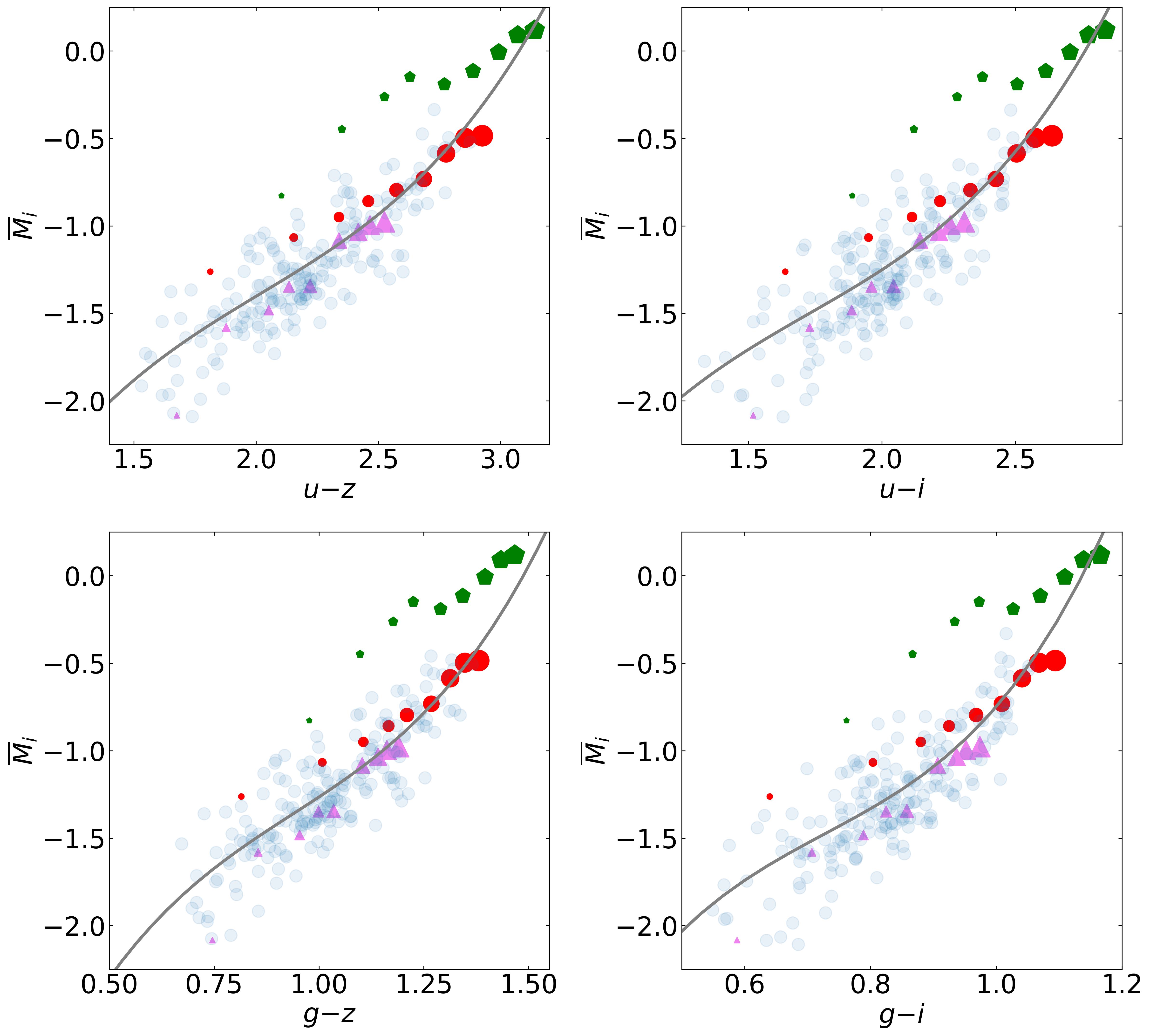} 
\caption{Comparison of the observed and model SBF-color relations. The gray lines show the third degree polynomial fits derived as described in the text. Shaded blue circles show the observational dataset used to obtain the calibration itself. The magenta, red and green symbols show the SPoT SSP models for three different metallicities (\feh=$-$0.35, 0.0 and $+$0.4, respectively). Ages in the range t=1-14 Gyr are shown with different symbol sizes (smaller for younger ages).
\label{spot}}
\end{center}
\end{figure*}

The comparison between empirical data and SSP model predictions is shown in Figure \ref{spot}. The observed galaxy colors generally match the sequences expected from the SSP models in the metallicity range ${-}0.35\leq \feh \leq {+}0.4$ dex, for ages between $1$ and $14$ Gyr. 
Each panel in the figure shows one of the SBF versus color relations derived in Section \S \ref{sec_cal}, together with the NGVS calibrating sample including the appropriate zero-point shifts as described above. For all the empirical relations shown (gray curves), we see a fair overlap with the entire age sequence of SSPs with $\feh=-0.35$ dex (violet triangles), with the solar metallicity SSP models older than $t{\,\sim\,}5$ Gyr, and with the oldest very metal-rich SSPs at $\feh=+0.4$ dex.

Interestingly, the empirical calibration for blue galaxies with $\Mbari\gtrsim-1$ mag matches well with the set of SSP models with $\feh=-0.35$~dex for the full age range from 1 to 14 Gyr. However, we do not expect that the dominant stellar component in the bluest sample galaxies is, in general, as young as 1~Gyr, since this this would likely be associated with an irregular galaxy morphology and/or the presence of dust that would complicate analysis of the SBF signal. As noted above, target galaxies were selected partly on the basis of having regular, ``clean" morphologies. Thus, these blue galaxies are likely to be somewhat older and more metal poor than the bluest models shown here.
Based on previous SPoT simulations, SSP models for older ages and $\feh<-0.5$ dex are also expected to cover the blue/bright side of the curves in Figure~\ref{spot} \citep[see, e.g.,][]{cantiello03}. New SSP models covering a wider range of [Fe/H] values and new photometric systems (like JWST, Euclid) are currently being developed by our team.

We emphasize that the comparisons shown here are based on SSP models derived from stellar evolution theory (e.g., stellar tracks, initial mass function, stellar atmospheres) combined with Monte-Carlo simulations. This demonstrates the robustness of the SBF technique as a distance indicator, in that it can be calibrated using purely theoretical arguments. It also highlights the usefulness of SBF measurements for studying the properties of unresolved stellar populations. We further discuss SBF as a stellar population tracer in \S\ref{sec_gradients}.

\section{Results and perspectives}
\label{sec_philosophy}

In this section, we use our new catalogue of SBF distances to examine the three-dimensional structure of the Virgo cluster. We include a discussion of possible SBF gradients in our program galaxies, and conclude with a summary of this work as well as some thoughts on areas for future research.

\subsection{Distances and Substructure}
\label{sec_distances}
Using the reference distances, $\dref$, reported in Table \ref{tab_measures}, we examine the distribution of the galaxies in our sample, both spatially and in the velocity-distance diagram. 
We restrict the analysis to only those galaxies with $q1$ and $q2$ quality codes.

\begin{figure*}
\begin{center}
\includegraphics[width=.85\textwidth]{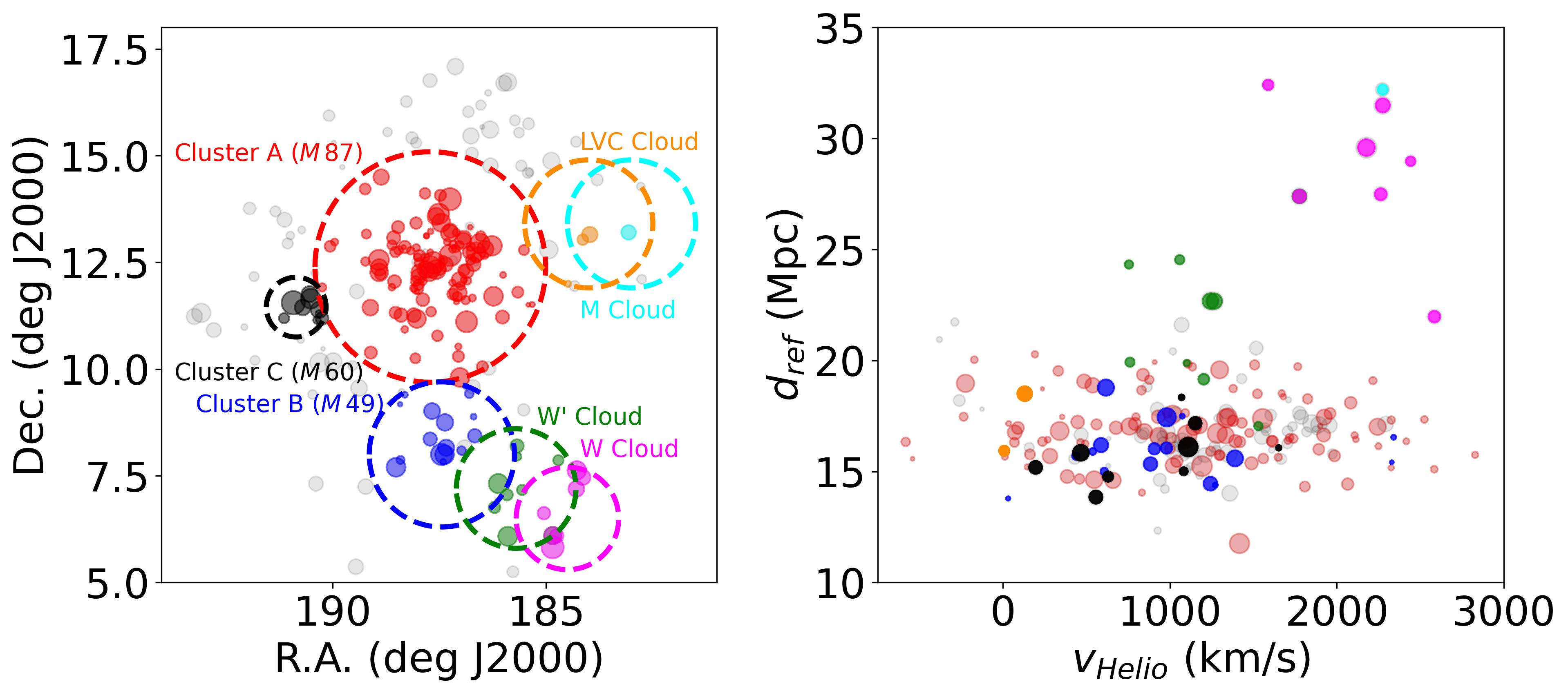} 
\caption{Left: map of the sky positions of the galaxies in our sample. Only class $q1$ and $q2$ galaxies in Table \ref{tab_measures} are plotted. The possible members of each substructure are identified through our $\dref$ measurements, recession velocities, and positions and radii (see Table \ref{tab_clouds}). The structures are color coded and labelled, with the size of the symbol scaling with galaxy magnitude. Note that the circles provide only rough indications of the structures, and in some cases represent upper values. Field galaxies, not associated with any of the substructure analysed, are indicated by the gray circles. Right: Hubble diagram for these same galaxies. Symbols and color coding are the same as in the left panel.
\label{2dplot}}
\end{center}
\end{figure*}

The left panel of Figure \ref{2dplot} shows position on the sky for our sample galaxies, with the outline of several Virgo cluster substructures identified in \citet{boselli14}. In the figure,  the half-virial radii for the A and B subclusters are from \citet{mclaughlin99} and \citet{ferrarese12}, while \citeauthor{boselli14} defined the other radii based on constraints to the angular distance from the overdensity peak.

For each substructure, we selected candidate member
galaxies using sky coordinates, distances, and recession velocities
with the following membership criteria. For all substructures, except
the M cloud and LVC, we: $i)$~selected all galaxies projected within
the substructure area, adopting the central positions and radii
reported in Table~\ref{tab_clouds} (Cols. 1-4); $ii)$ for the selected sources, 
we derived the median and the standard deviation (from the median
absolute deviation) of the distances and the recession velocities
reported in Table~\ref{tab_measures}; $iii)$ rejected from the selected sample all
sources that are more than 3-$\sigma$ outliers with respect to the mean
distance and velocity. For the distant M cloud (where we can identify
only a single member in our sample), in addition to angular projection
proximity, we required $\dref>25$~Mpc. For membership in the LVC, we
required projection proximity and $v_{h}<800$ \kms\ \citep{boselli14}.

The substructures and their associated galaxies are color coded in the left panel of Figure~\ref{2dplot}, as labeled. Sample galaxies that fall outside the boundaries of all substructures (``Virgo field") are represented by light-gray circles. The right panel of the figure shows a Hubble diagram of the substructures including Virgo field galaxies, using the same color coding as in the left panel. As in Paper~I, there is no obvious distinction in the Hubble diagram among the subclusters around M\,87, M\,49 and M\,60, although both the W and \Wprime\ clouds do appear distinct from the main cluster.  

\begin{figure*}
\begin{center}
\includegraphics[width=.85\textwidth]{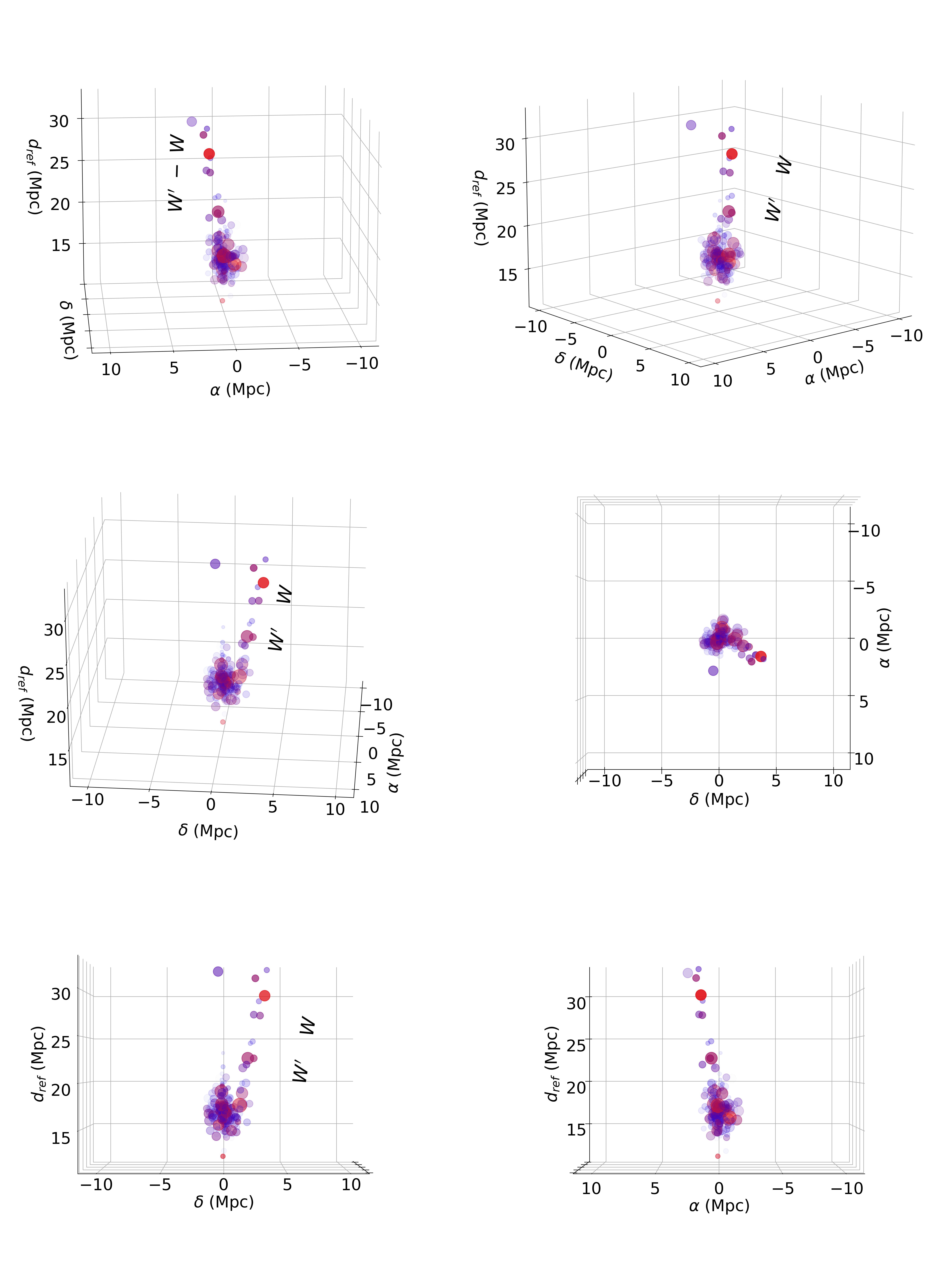} 
\caption{The three-dimensional distribution of the sample of galaxies in
Figure~\ref{2dplot}, shown here from different vantage points.
The symbol size scales with galaxy luminosity, and symbol color is coded
according to the actual \gz\ color. 
The locations of the \Wprime\ and W clouds are shown in several panels.
Additional representations of the three-dimensional distribution are shown in the Appendix.
\label{3dplot}}
\end{center}
\end{figure*}

The transitional region between the main Virgo cluster and the \Wprime/W clouds is better appreciated in the three-dimensional plots shown in Figure \ref{3dplot}. In this figure, we see a sequence of about a dozen galaxies lying along a `filament' from the core of subcluster~B toward the W cloud, and covering the \Wprime~region. This feature may  be a projection resulting from sampling effects of the so-called W-M sheet \citep[see][and references therein]{kim16,castignani22}.

The main properties of the substructures inspected here are summarized in Table \ref{tab_clouds}, which contains the substructure identifier, coordinates, radius from \citet{boselli14}, the number of galaxies in our sample that we identify as members ($N_\mathrm{str}$), median velocity, median distance, and estimated $\rms$ depth after correcting for expected scatter from distance errors (with the bootstrap uncertainty given in parentheses).
Although we formally calculate a depth of ${\sim\,}4$ Mpc for the \Wprime\ cloud, 
we note that this structure is prone to contamination from both the richer foreground subcluster~B and
the background W cloud. It may be that a more useful physical definition of \Wprime\ would be the giant
elliptical NGC\,4365 at ${\sim\,}23$ Mpc and its close satellites. In that case, the group would be a
much more compact knot within the filament-like W cloud extending out to beyond 30 Mpc.

\begin{deluxetable*}{lrrcrrcc}
\label{tab_clouds}
\tablecaption{Main characteristics of Virgo cluster substructures.} 
\tablewidth{0pt}
\tablehead{
\colhead{Structure} & \colhead{R.A.} & \colhead{Dec.} & \colhead{Radius} & \colhead{$N_\mathrm{str}$} & \colhead{$\langle v_h \rangle_\mathrm{med}$} & \colhead{$\langle d \rangle_\mathrm{med}$} & \colhead{Depth} \\
 & (J2000) & (J2000) & (deg) &  & (\kms) & (Mpc) & (Mpc)} \startdata
Cluster A (M\,87) & 187.7 & 12.4 & 2.7 & 113 (283/1503) & 1157 & 16.8 & 0.6 ($\pm0.1$)\\
Cluster B (M\,49) & 187.4 &  8.0 & 1.7 &  16 (97/425)   & 944 & 15.8 & 0.4 ($\pm0.3$)\\
Cluster C (M\,60) & 190.9 & 11.4 & 0.7 &   9 (25/79)    & 1070 & 15.8 & 1.0 ($\pm0.4$)\\
\Wprime~cloud     & 185.7 &  7.2 & 1.4 &   9 (75/298)   & 1203 & 22.7 & 4.1 ($\pm0.8$)\\
W cloud           & 184.5 &  6.5 & 1.2 &   7 (51/136)   & 2263 & 29.0 & 1.4 ($\pm1.3$)\\
LV cloud          & 184.0 & 13.4 & 1.5 &   3 (180/232)  & 70 & 15.9 & \nodata \\
M cloud           & 183.0 & 13.4 & 1.5 &   1 (34/163)   & 2273 & 32.2  & \nodata 
\enddata
\end{deluxetable*}

As previously mentioned, the SBF analysis favors galaxies with smooth
surface brightness profiles. This preference may result in
early-type galaxies, with smoother profiles, being measured across a wider 
range of distances than later type galaxies. Additionally,
smooth early-type galaxies tend to reside in the denser regions of a cluster
\citep[e.g.,][]{dressler1980}. The combination of these factors implies that, at a fixed image
depth and quality, more reliable SBF distances would be measured
toward the densest substructures.
Hence, the properties of the structures reported in Table \ref{tab_clouds},
especially the depth, should be interpreted in light of this bias.

Table \ref{tab_clouds} also reports the number of galaxies detected in the NGVS
survey around the projected area of each subcluster region using the 
catalogs from \citet{ferrarese20} and \citet{ferrarese24},
adopting the same redshift properties used to identify galaxies
in the substructure, as well as the numbers without any redshift
constraint (these numbers are given in parentheses in the $N_{str}$ column). 
While it would be possible to incorporate additional Virgo galaxy
distances from other methods \citep[e.g.,][]{gavazzi1999}, as noted 
in Section \ref{sec_cal}, the combined heterogeneity of literature samples
and the typically much larger scatter from distances derived using 
alternative indicators (e.g., both the Tully-Fisher and fundamental plane methods
have distance uncertainties $\sim 20\%$, larger than the sizes of 
most of the structures we have discussed), as compared to the homogeneity of 
the NGVS sample and the small intrinsic scatter of SBF distances (Sec.~\ref{sec_cal}),
discourage us from combining our measurements with literature data for this analysis.

\subsection{A Closer Look at Cluster~A}
\label{doubleA}

\begin{figure*}
\begin{center}
\includegraphics[width=.87\textwidth]{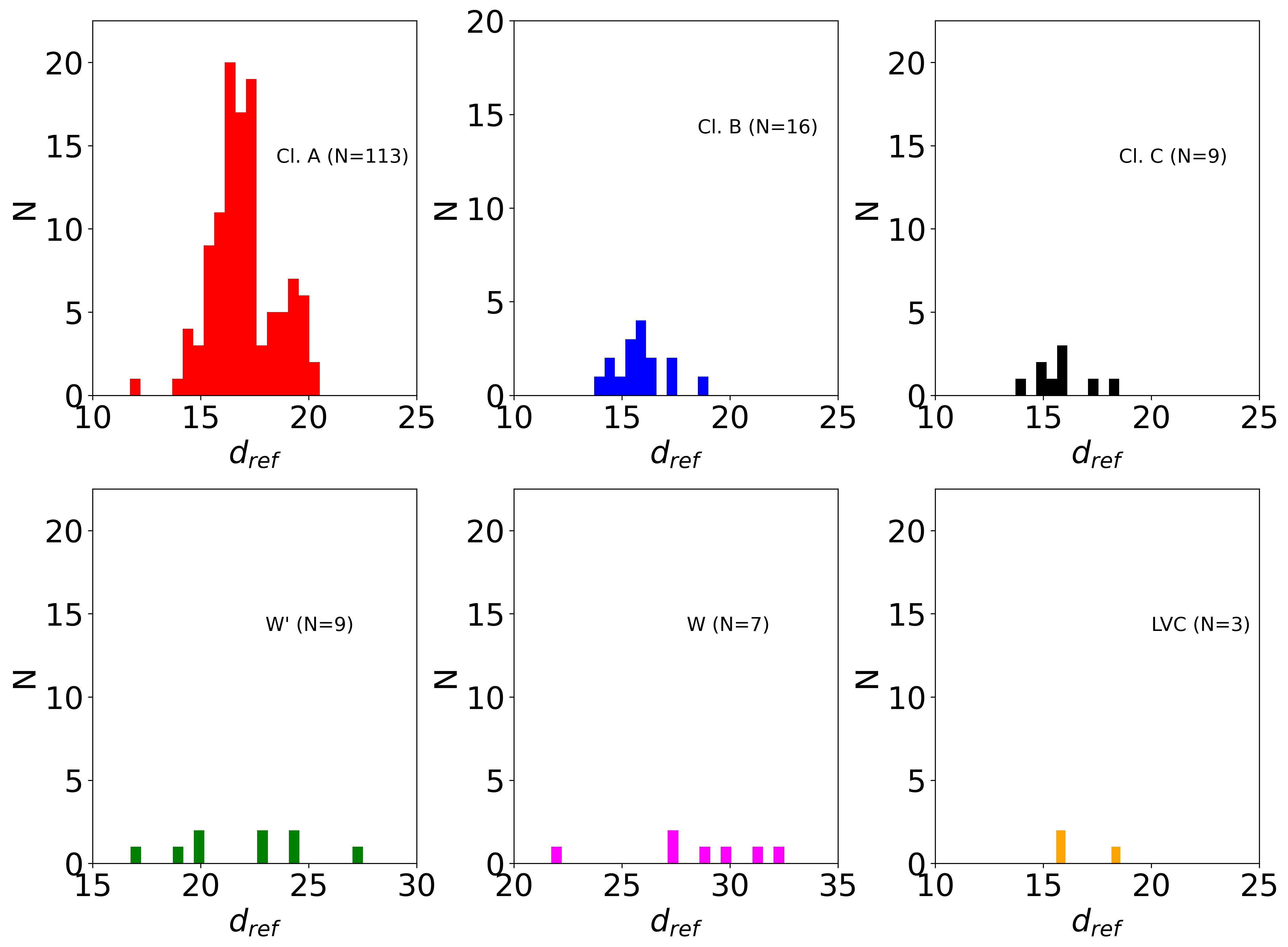} 
\caption{Histograms of galaxy distances within each of the substructures listed in Table~\ref{tab_clouds}. Subcluster~A (top left panel) shows an apparent bimodality in its distance distribution with distinct peaks at $16.5\pm0.1$ and $19.4\pm0.2$ Mpc.
\label{dhistos}}
\end{center}
\end{figure*}

In examining the spatial distribution of the galaxies in our sample, we noticed an apparent gap in distance near 18~Mpc. The feature is most pronounced in the distribution of distances for galaxies grouped into the A~subcluster. Figure~\ref{dhistos} shows the histograms for all of the identified structures with more than one member. The bimodality in the histogram of subcluster~A is clear; we used the Gaussian Mixture Modeling (GMM) code of \citet{muratov10} to quantify this impression. The GMM code compares the goodness of fit for single and double Gaussian models of the unbinned distribution, then uses bootstrap resampling to assess the uncertainties. In the case of subcluster~A, it finds that the distances prefer a double Gaussian model with $\sim\,$99.9\% confidence.  For the preferred double Gaussian model, the two peaks occur at $16.5\pm0.1$ and $19.4\pm0.2$ Mpc, with dispersions (uncorrected for measurement error) of $1.0\pm0.1$ and $0.5\pm0.1$~Mpc, respectively. According to GMM, the nearer peak comprises 82\% of the sample. Based on the appearance of the histogram, the strong preference for two Gaussians is not surprising, but there is, of course, no reason to expect that distances should follow a Gaussian distribution.

A more meaningful question is: how significant is the gap in the distance distribution? In an attempt to address this, GMM calculates the separation of the fitted peaks in units of the combined dispersion, and assesses the significance of the gap based on this statistic. Using this approach, GMM finds the gap is significant with 90\% confidence. However, there is again an underlying assumption of Gaussianity in this analysis. Finally, the software package provided by \citet{muratov10} also includes code to calculate the ``dip" statistic of \citet{hartigan85}, which provides a measure of the significance of multimodality based on the maximum difference of the sorted empirical distribution with respect to the nonparametric unimodal distribution that minimizes this maximum difference. According to this test, we cannot rule out a unimodal distance distribution for subcluster~A at even $1\sigma$ significance. \citet{muratov10} find a similar result for the distribution of metallicities in Galactic globular clusters, and conclude the dip test is ``less powerful" than GMM. However, it is robust against the assumption of Gaussianity.

In any case, the distance distribution of galaxies in subcluster~A is certainly suggestive of bimodality. 
We performed tests with alternative distance calibrations for \Mibar, such as using \gi\ instead of \uz\ for the stellar population dependence, and adopting a linear rather than a polynomial relation. However, the apparent bimodality remains, with similarly high significance $\gta99.8\%$ from GMM, and low significance
$\lta50\%$ from the dip test. Thus, the double-peaked structure is not the result of the adopted calibration relation.

\begin{figure*}
\begin{center}
\includegraphics[width=.85\textwidth]{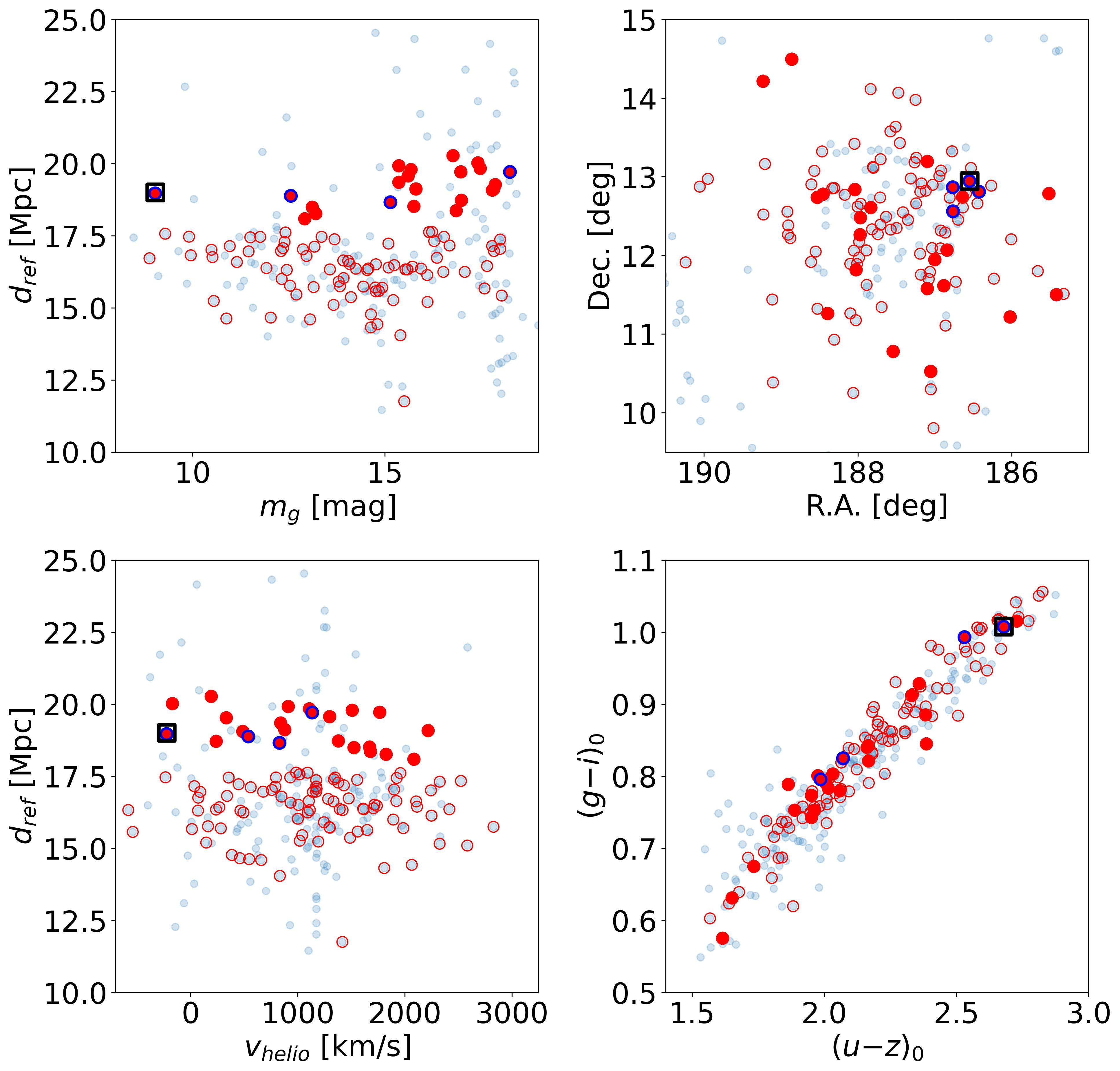} 
\caption{A closer look at subcluster~A. Plots of distance versus magnitude, right ascension and declination, distance versus velocity, and \gi\ versus \uz\ color for galaxies in the main $d{\,\sim\,}16.5$ Mpc distance peak (open red circles)
and the smaller $d{\,\sim\,}19.4$ Mpc distance peak (filled red circles) of subcluster~A, where the separation is taken as 18~Mpc. Other Virgo galaxies in our sample are shown as smaller gray points.  The giant elliptical M\,86 (VCC\,881) is marked by a  black square, while three galaxies near it on the sky and with similar velocities are marked by blue circles. Apart from these few galaxies, there is no obvious grouping in position and velocity of the secondary peak galaxies around M\,86.  The two distance subgroups fall along the same color-color relation, so the putative bimodality does not appear to be a consequence
of stellar population differences.
\label{clusterAplots}}
\end{center}
\end{figure*}

The most luminous galaxy in the secondary distance peak is M\,86 (VCC 881) at $19.1\pm1.1$ Mpc and with a velocity $v_h = -224$ \kms. While there are a few other galaxies (VCC 828, 997, and 1414) with  distances of 19~to 20~Mpc, $v_h\lesssim500$ \kms, and projected locations within 0.5~Mpc of M\,86, there is no obvious spatial concentration of the galaxies in the secondary peak around this bright galaxy. Figure~\ref{clusterAplots} compares the distances, magnitudes, sky locations, velocities, and colors of galaxies in the primary and secondary distance peaks of subcluster~A, where we use 18~Mpc as the dividing line. M\,86 is marked with a square box in the figure. Most of the galaxies in the secondary peak are dwarfs with $g$ magnitude $m_g>15$~mag, but overall the galaxies are not distinguished from the rest of the subcluster~A galaxies by any parameter other than distance. We conclude that the apparent bimodality in the distances within subcluster~A is intriguing, but needs confirmation and further exploration with an even larger sample of high-quality distances.

\subsection{SBF versus color gradients}
\label{sec_gradients}

In addition to the main annulus used for our SBF analysis, for every galaxy in our sample, we measured SBF and colors in three to five smaller and distinct circular concentric annuli, to analyze the radial behaviour of integrated stellar population properties using combined SBF and color gradients. In examining these radial gradients, we limited our sample to galaxies that have either $q1$ or $q2$ quality flags, and for which the SBF and color gradients are monotonic. 

\begin{figure*}
\begin{center}
\includegraphics[width=.85\textwidth]{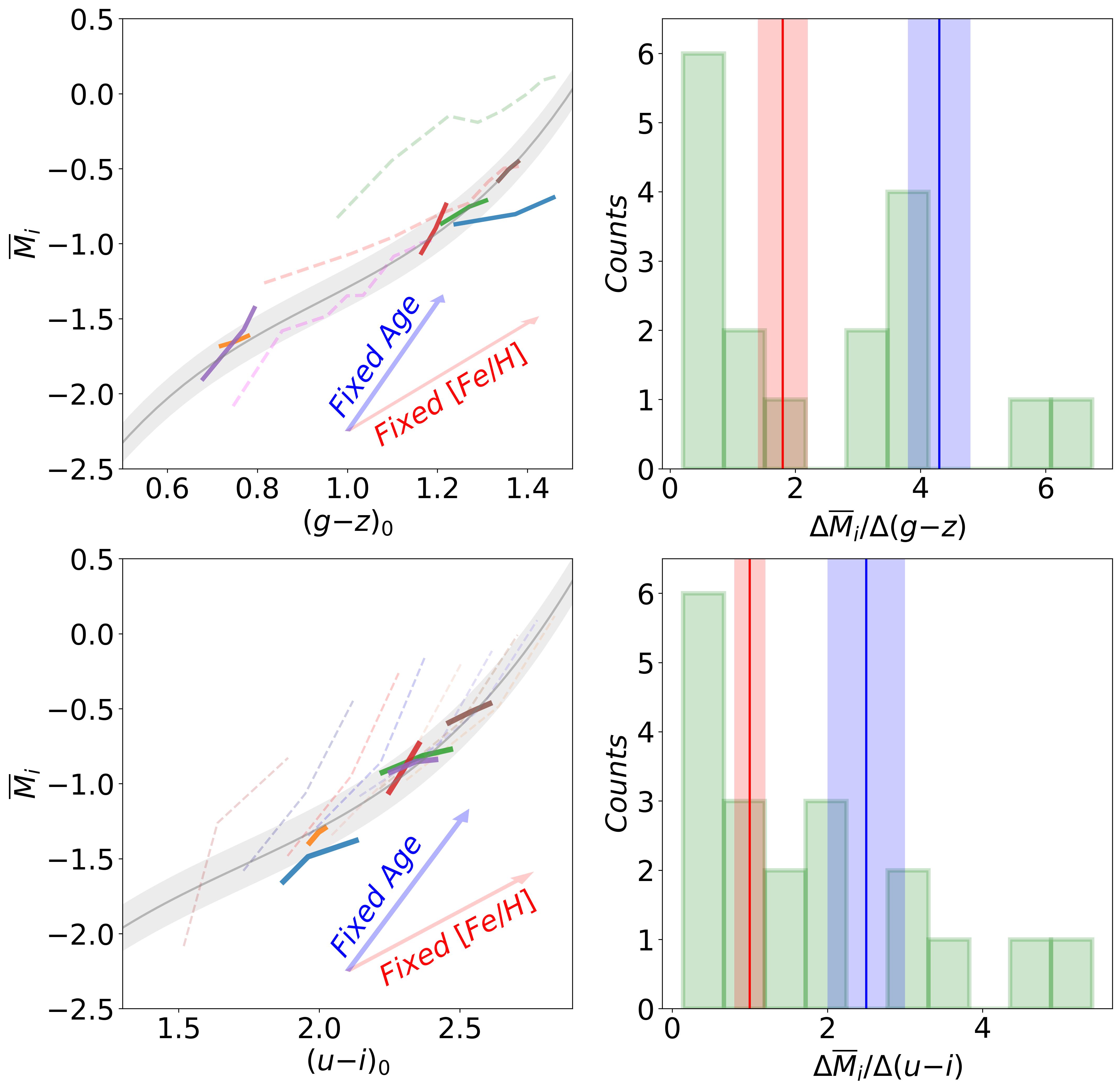} 
\caption{Left panels: \Mibar\ versus color diagrams for \gz\ (upper panel) and \ui\ (lower panel). The gray line and area show the fits of Table \ref{tab_fit} and the scatter. The dashed lines mark the position of SPoT SSP models: connected by \feh\ homogeneity (upper panel) or by age homogeneity (lower panel). The thick coloured lines show the SBF-color sequence for a selection of galaxies. The blue (red) arrow indicates the mean slope for SBF versus color derived from SSP models with varying \feh (age) at fixed age (metallicity). The direction of the arrows indicates the increase of Age or \feh. 
Right panels: histograms of the slopes of the SBF-color relations for the selected sample. The shaded blue and red areas and lines mark the position of the mean slopes derived from the SPoT SSP models for pure \feh~ and for pure age variations, respectively.
\label{allgrads}}
\end{center}
\end{figure*}

The left panels of Figure \ref{allgrads} show SBF amplitudes versus color data for the galaxies where the uncertainty on the fitted slope, $\Delta \Mibar/\Delta color$, is smaller than one half of the expected {\it pure age} gradient, as explained below. Using the SPoT SSP models in Table \ref{tab_models}, we estimated the $\Mibar$ versus color gradient as follows: we first fix the metallicity, \feh , and, for the entire set of ages available, derive the slope of the \Mibar-color relation for each of the three \feh\ available. The dashed lines in the upper left panel of Figure~\ref{allgrads} indicate sequences in age at fixed \feh. For the two colors considered, the mean slopes derived are:
$$\frac{\delta \Mibar}{\delta \gz} \Bigr\rvert_{\feh}=1.8\pm0.4\,,$$
$$\frac{\delta \Mibar}{\delta \ui} \Bigr\rvert_{\feh}=1.0\pm 0.2\,.$$

Similarly, we derived the slopes of $\Mibar$ versus color at fixed ages. In the lower left panel of Figure \ref{allgrads}, the dashed lines connect the predictions from the SPoT SSP models for coeval (fixed age) populations with differing values of \feh.
The median slopes of the fitted linear equations are:
$$\frac{\delta \Mibar}{\delta \gz} \Bigr\rvert_{Age}=4.3\pm0.5\,,$$
$$\frac{\delta \Mibar}{\delta \ui} \Bigr\rvert_{Age}=2.5\pm 0.5\,.$$


Rather than the exact values obtained from SSP models, what we highlight here is the evidence that SBF versus color trends driven by \feh variations at fixed age are predicted to have a slope two times steeper than those at fixed metallicity \feh. In the left panels of Figure \ref{allgrads}, we illustrate the direction of age gradients (indicated by the red arrow labeled ``Fixed \feh") and metallicity gradients (shown by the blue arrow labeled ``Fixed Age").

In the right panels of Figure \ref{allgrads}, we plot the histograms of the SBF-color slopes derived for all class $q1$ and $q2$ galaxies with monotonic  radial gradients in SBF magnitude and color, without the limitation on maximum error of the slope (adopted in the left panels for sake of clarity). In the panels, we represent with green histograms the sample of $\sim$40 targets thus selected and the mean slopes derived from SSP models, with the same color coding as in the left panels.

From the histograms of gradients for both colors, we find that the median for the observed data lies between the two fiducial lines, with a slight preference for smaller slopes (i.e., age-driven, fixed \feh) compared to larger slopes (fixed Age, \feh-driven). 
Additionally, we assessed the gradient differences between bright and faint galaxies, using a threshold at $B_T=13$ mag. Fainter galaxies tend to exhibit slightly larger gradients,
although the statistical significance is weak.

Of course, most galaxies likely have both age and \feh\ gradients to some degree.
Nevertheless, it is interesting to note that in the lower-left panel of Figure \ref{allgrads}, 
most of the multi-annuli data (thick broken lines with different colors) have a $\Delta \Mibar/\Delta \ui<2$, indicative of significant age gradients. The biggest exception is VCC\,1720 (dark-red line, at \ui$\sim2.3$ mag and \Mibar$\sim-0.9$ mag), which appears to align with the model sequences at fixed age, with a $\Delta \Mibar/\Delta \ui=3.1\pm0.1$. For the same galaxy, we find  $\Delta \Mibar/\Delta \gz=5.9\pm0.5$, more than twice as steep as the median slope for the sample in the upper-right panel.

\subsection{Perspectives}

Taking advantage of the depth and image quality of the NGVS dataset, we have presented SBF measurements and distances for 278 galaxies within the 104~deg$^2$ survey footprint. 
This sample of distances is several times larger than in any previous SBF study of the Virgo region.
These results highlight the great potential of the SBF method for deriving high-quality distances for large sample of galaxies in upcoming wide-field optical and near-IR imaging surveys.

 With a 5-$\sigma$ point-source depth of $\sim25.8$ mag and $i$-band seeing $\leq0\farcs7$, the NGVS data are about one magnitude shallower than what is expected from Rubin/LSST. Since the distance to which SBF can be reliably measured is often limited (for bright galaxies) by the ability to remove contaminating sources, the greater depth of the LSST data implies a distance limit potentially several times larger than in this study, for which the maximum measured distance is $\sim$33 Mpc. Similarly, the Euclid satellite, now in operation, will have an $H$-band point-source limit of $\sim 24$ AB mag, with an (undersampled) PSF of ${\sim\,}0\farcs3$. This combination of angular resolution, depth, sky coverage, and the benefit of minimal impact from dust, will be ideal for measuring SBF distances for large samples of nearby galaxies, complementing those measured in the optical with Rubin/LSST.

On slightly longer timescales, the Nancy Grace Roman Space Telescope 
will be of enormous interest for SBF measurements. Roman will have the same aperture as the Hubble Space Telescope and similar resolution, but a $\sim100$ times larger field of view and superior sensitivity at infrared wavelengths. With a $5\sigma$ point-source depth of 28~mag in 1 hr in the $J$ and $H$ bands, Roman should deliver phenomenal survey depth and breadth, making it the ultimate machine for churning out SBF distances in abundance
\citep[e.g.,][]{greco21,blakeslee23}.
Notably, all three of the above-mentioned facilities will observe the sky in multiple passbands, making it possible to derive exquisite calibrations for the SBF distances, as well as to measure SBF colors (i.e., differences of SBF magnitudes in multiple bands), a unique tool for probing stellar populations.
%

To realize the full benefit of these deep, wide-area, multi-band surveys,
a new paradigm for SBF measurements will be essential. Efficient and robust automated procedures must be developed for measuring SBF amplitudes and distances in large numbers of galaxies without requiring extensive human intervention. 
We are presently developing such automated SBF pipelines, taking advantage of the large, well-tested sample of SBF measurements that we have presented here based on the multi-band NGVS observations.
 
Finally, in addition to the potential of wide-area surveys for measuring unprecedented numbers
of SBF distances across the sky, the large apertures, high near-infrared sensitivities, and excellent spatial resolution of JWST and future AO-assisted instruments on ELTs, will allow the SBF method to be pushed to unprecedented distances for targeted observations of individual galaxies. Once calibrated
from the TRGB method, this will provide a fully independent distance ladder
rivaling the more familiar one based on Cepheids and Type~Ia supernovae.
Altogether, the outlook for the SBF technique as a distance and stellar population indicator is excellent.

\vspace{4mm}

\begin{acknowledgments}
M.C.\ and G.R.\ acknowledge INAF for supporting the project {\it Gravitational Wave Astronomy with the first detections of LIGO and Virgo experiments} (GRAWITA, PI: Enzo Brocato). M.C. gratefully acknowledges the invaluable professional support provided by Sabrina Ciprietti.
This paper is based on observations obtained with Mega-Prime/MegaCam, a joint project of CFHT and CEA/IRFU, at the Canada–France–Hawaii Telescope (CFHT), which is operated by the National Research Council (NRC) of Canada, the Institut National des Sciences de l’Univers of the Centre National de la Recherche Scientifique (CNRS) of France, and the University of Hawaii. This research used the facilities of the Canadian Astronomy Data Centre operated by the National Research Council of Canada with the support of the Canadian Space Agency.

This research has made use of the NASA/IPAC Extragalactic Database (NED), which is funded by the National Aeronautics and Space Administration and operated by the California Institute of Technology.

\end{acknowledgments}

 \software{Astropy \citep{astropy:2013,astropy:2018},           
          Source Extractor \citep{bertin96},
          TOPCAT \citep{taylor05}}



\appendix

\section{Additional Plots of the Three-Dimensional  Distribution of the Sample}

In this Appendix, we present additional plots illustrating the three-dimensional distribution of galaxies in the Virgo cluster.

\begin{figure*}
\begin{center}
\includegraphics[width=.85\textwidth]{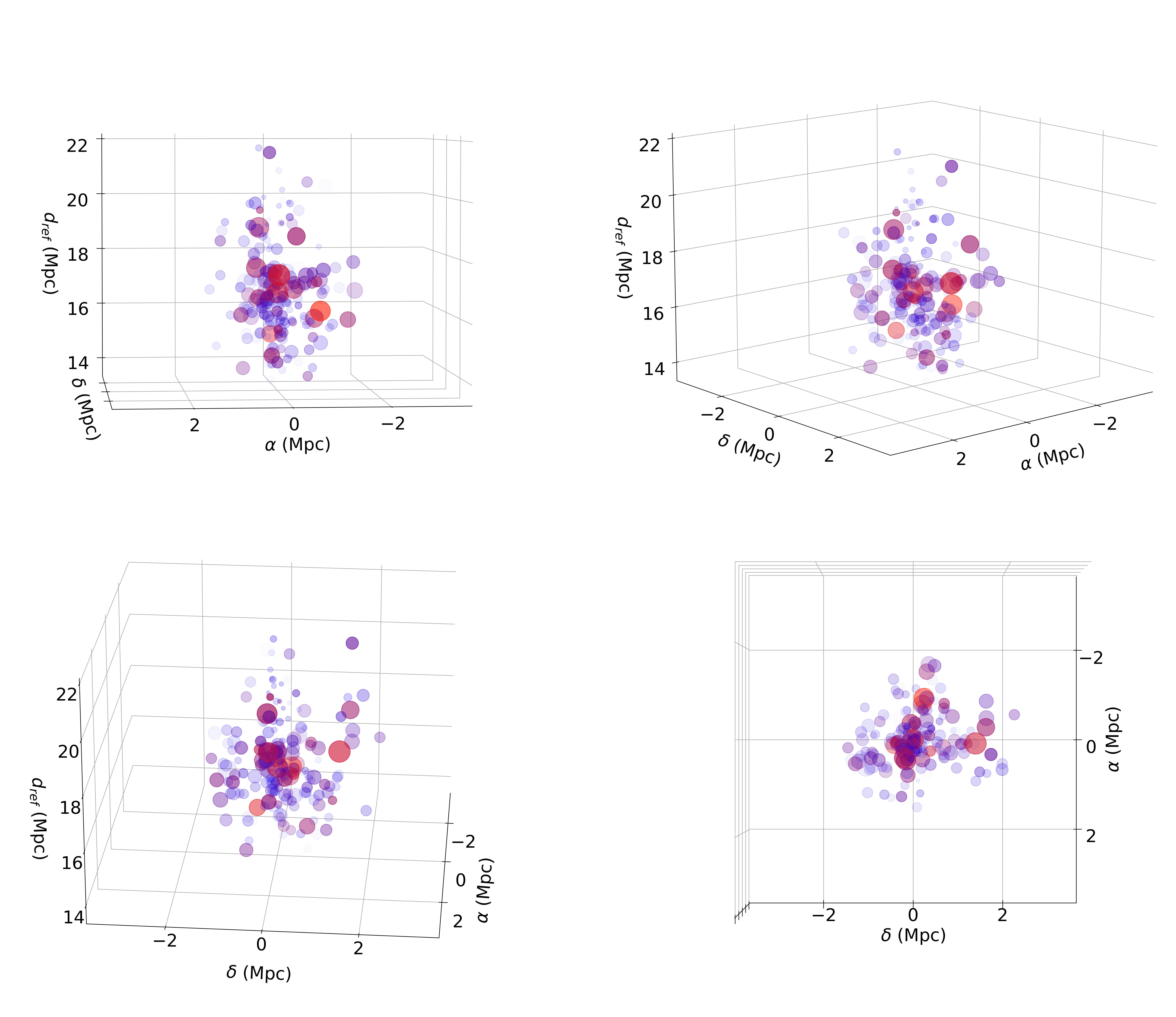} 
\caption{Same as in Figure \ref{3dplot}, but limited to the sample of galaxies $12.5\leq d (Mpc) \leq 22$. The distinctive structure of the cluster, elongated along the line of sight, is clearly evident in this representation.
\label{3dplot_zoom}}
\end{center}
\end{figure*}


\begin{figure*}
\begin{center}
\includegraphics[width=.85\textwidth]{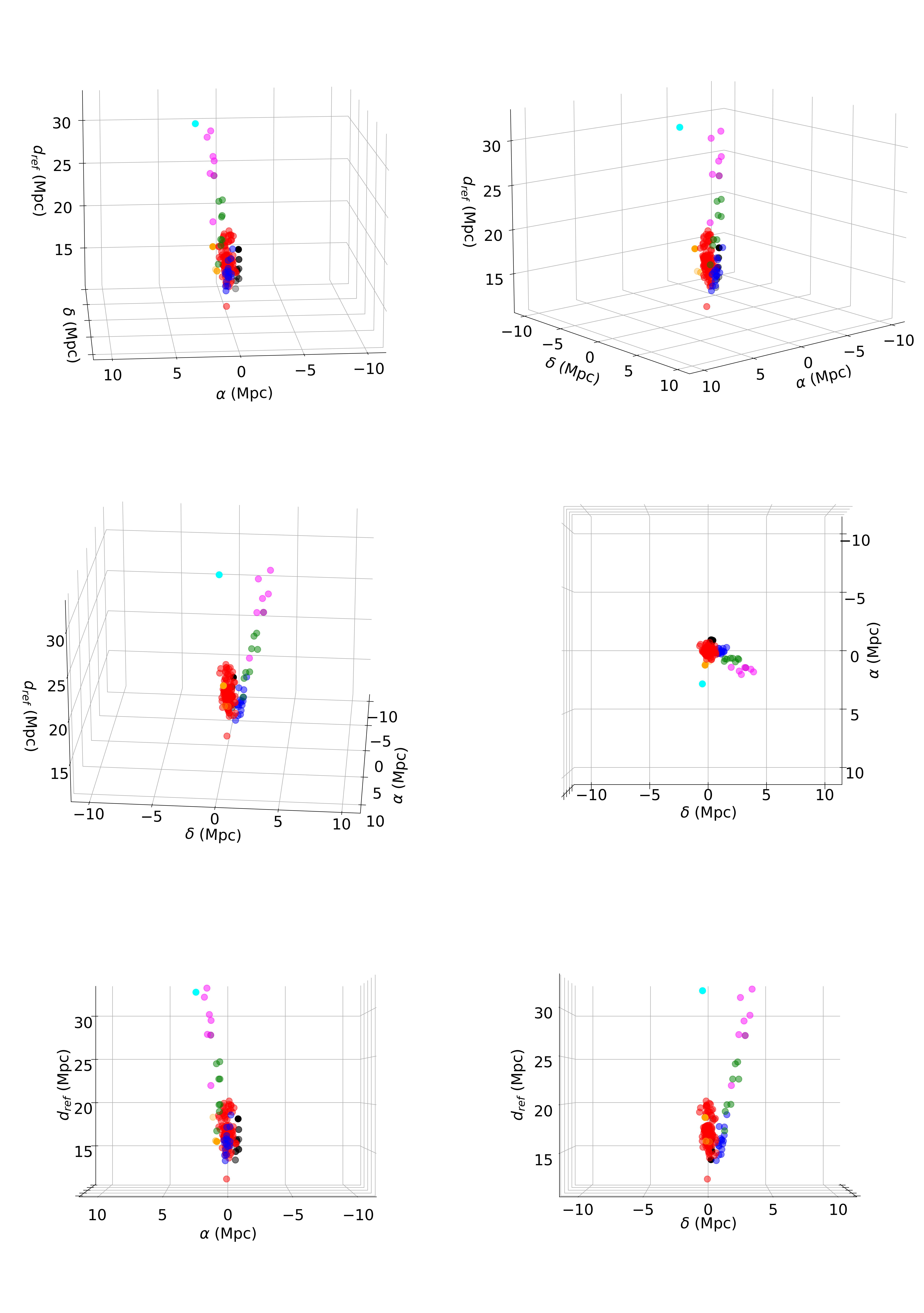} 
\caption{3D distribution of the galaxies associated with the subclusters, groups, and clouds using the same color-coding adopted in Figure \ref{2dplot}. 
In this view, a filamentary structure appears to connect Cluster B and the \Wprime and W clouds.
\label{3dplot_clouds}}
\end{center}
\end{figure*}

\begin{figure*}
\begin{center}
\includegraphics[width=.85\textwidth]{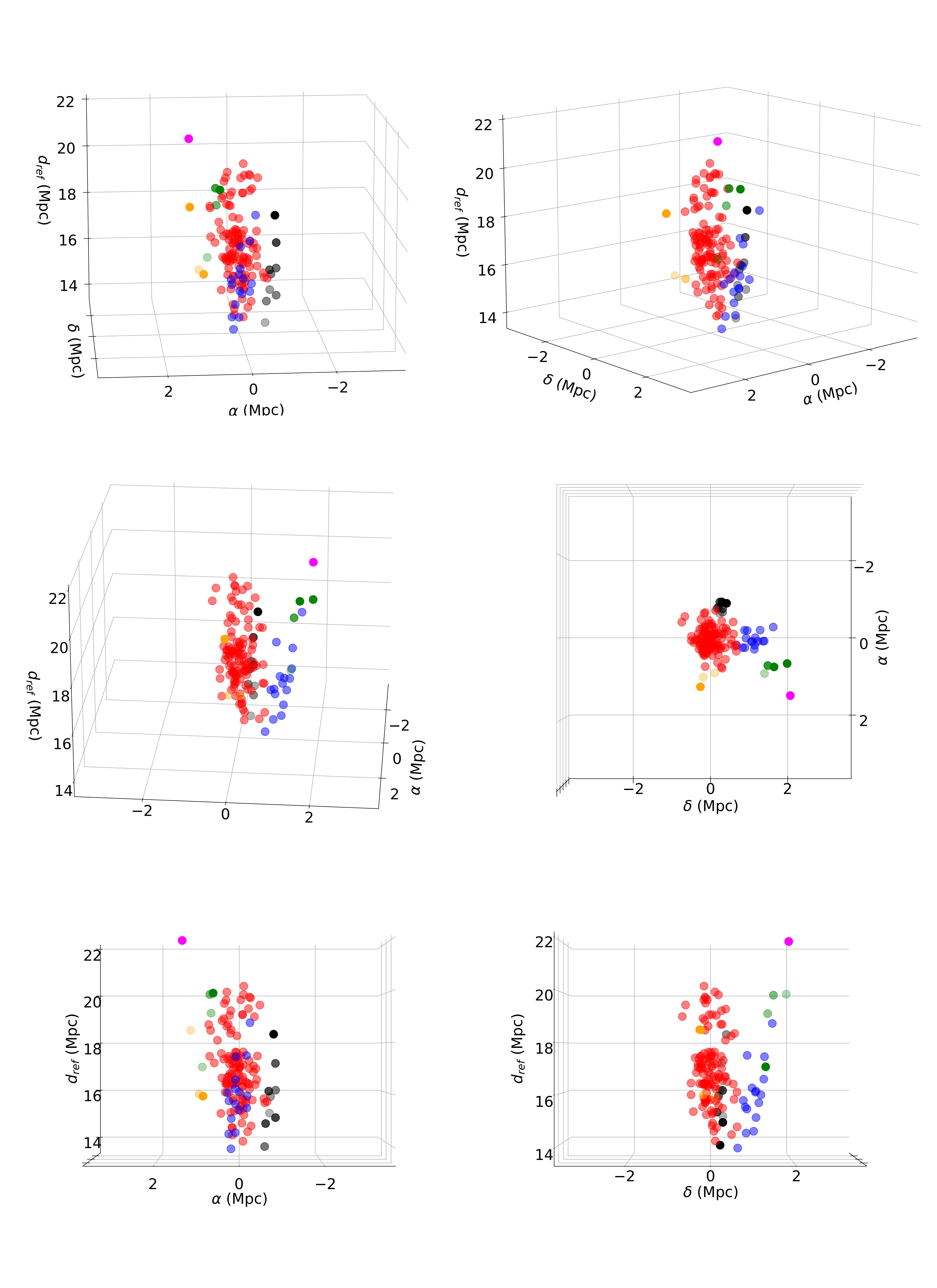} 
\caption{Same as in Figure \ref{3dplot_clouds} but limited to the sample of galaxies $12.5\leq d (Mpc) \leq 22$.
\label{3dplot_clouds_zoom}}
\end{center}
\end{figure*}


\bibliography{cantiello_ngvs_II}{}

\begin{thebibliography}{}
\expandafter\ifx\csname natexlab\endcsname\relax\def\natexlab#1{#1}\fi
\providecommand{\url}[1]{\href{#1}{#1}}
\providecommand{\dodoi}[1]{doi:~\href{http://doi.org/#1}{\nolinkurl{#1}}}
\providecommand{\doeprint}[1]{\href{http://ascl.net/#1}{\nolinkurl{http://ascl.net/#1}}}
\providecommand{\doarXiv}[1]{\href{https://arxiv.org/abs/#1}{\nolinkurl{https://arxiv.org/abs/#1}}}

\bibitem[{{Astropy Collaboration} {et~al.}(2013){Astropy Collaboration},
  {Robitaille}, {Tollerud}, {Greenfield}, {Droettboom}, {Bray}, {Aldcroft},
  {Davis}, {Ginsburg}, {Price-Whelan}, {Kerzendorf}, {Conley}, {Crighton},
  {Barbary}, {Muna}, {Ferguson}, {Grollier}, {Parikh}, {Nair}, {Unther},
  {Deil}, {Woillez}, {Conseil}, {Kramer}, {Turner}, {Singer}, {Fox}, {Weaver},
  {Zabalza}, {Edwards}, {Azalee Bostroem}, {Burke}, {Casey}, {Crawford},
  {Dencheva}, {Ely}, {Jenness}, {Labrie}, {Lim}, {Pierfederici}, {Pontzen},
  {Ptak}, {Refsdal}, {Servillat}, \& {Streicher}}]{astropy:2013}
{Astropy Collaboration}, {Robitaille}, T.~P., {Tollerud}, E.~J., {et~al.} 2013,
  \aap, 558, A33, \dodoi{10.1051/0004-6361/201322068}

\bibitem[{{Astropy Collaboration} {et~al.}(2018){Astropy Collaboration},
  {Price-Whelan}, {Sip{\H{o}}cz}, {G{\"u}nther}, {Lim}, {Crawford}, {Conseil},
  {Shupe}, {Craig}, {Dencheva}, {Ginsburg}, {Vand erPlas}, {Bradley},
  {P{\'e}rez-Su{\'a}rez}, {de Val-Borro}, {Aldcroft}, {Cruz}, {Robitaille},
  {Tollerud}, {Ardelean}, {Babej}, {Bach}, {Bachetti}, {Bakanov}, {Bamford},
  {Barentsen}, {Barmby}, {Baumbach}, {Berry}, {Biscani}, {Boquien}, {Bostroem},
  {Bouma}, {Brammer}, {Bray}, {Breytenbach}, {Buddelmeijer}, {Burke},
  {Calderone}, {Cano Rodr{\'\i}guez}, {Cara}, {Cardoso}, {Cheedella}, {Copin},
  {Corrales}, {Crichton}, {D'Avella}, {Deil}, {Depagne}, {Dietrich}, {Donath},
  {Droettboom}, {Earl}, {Erben}, {Fabbro}, {Ferreira}, {Finethy}, {Fox},
  {Garrison}, {Gibbons}, {Goldstein}, {Gommers}, {Greco}, {Greenfield},
  {Groener}, {Grollier}, {Hagen}, {Hirst}, {Homeier}, {Horton}, {Hosseinzadeh},
  {Hu}, {Hunkeler}, {Ivezi{\'c}}, {Jain}, {Jenness}, {Kanarek}, {Kendrew},
  {Kern}, {Kerzendorf}, {Khvalko}, {King}, {Kirkby}, {Kulkarni}, {Kumar},
  {Lee}, {Lenz}, {Littlefair}, {Ma}, {Macleod}, {Mastropietro}, {McCully},
  {Montagnac}, {Morris}, {Mueller}, {Mumford}, {Muna}, {Murphy}, {Nelson},
  {Nguyen}, {Ninan}, {N{\"o}the}, {Ogaz}, {Oh}, {Parejko}, {Parley}, {Pascual},
  {Patil}, {Patil}, {Plunkett}, {Prochaska}, {Rastogi}, {Reddy Janga},
  {Sabater}, {Sakurikar}, {Seifert}, {Sherbert}, {Sherwood-Taylor}, {Shih},
  {Sick}, {Silbiger}, {Singanamalla}, {Singer}, {Sladen}, {Sooley},
  {Sornarajah}, {Streicher}, {Teuben}, {Thomas}, {Tremblay}, {Turner},
  {Terr{\'o}n}, {van Kerkwijk}, {de la Vega}, {Watkins}, {Weaver}, {Whitmore},
  {Woillez}, {Zabalza}, \& {Astropy Contributors}}]{astropy:2018}
{Astropy Collaboration}, {Price-Whelan}, A.~M., {Sip{\H{o}}cz}, B.~M., {et~al.}
  2018, \aj, 156, 123, \dodoi{10.3847/1538-3881/aabc4f}

\bibitem[{{Astropy Collaboration} {et~al.}(2022){Astropy Collaboration},
  {Price-Whelan}, {Lim}, {Earl}, {Starkman}, {Bradley}, {Shupe}, {Patil},
  {Corrales}, {Brasseur}, {N{"o}the}, {Donath}, {Tollerud}, {Morris},
  {Ginsburg}, {Vaher}, {Weaver}, {Tocknell}, {Jamieson}, {van Kerkwijk},
  {Robitaille}, {Merry}, {Bachetti}, {G{"u}nther}, {Aldcroft},
  {Alvarado-Montes}, {Archibald}, {B{'o}di}, {Bapat}, {Barentsen}, {Baz{'a}n},
  {Biswas}, {Boquien}, {Burke}, {Cara}, {Cara}, {Conroy}, {Conseil}, {Craig},
  {Cross}, {Cruz}, {D'Eugenio}, {Dencheva}, {Devillepoix}, {Dietrich},
  {Eigenbrot}, {Erben}, {Ferreira}, {Foreman-Mackey}, {Fox}, {Freij}, {Garg},
  {Geda}, {Glattly}, {Gondhalekar}, {Gordon}, {Grant}, {Greenfield}, {Groener},
  {Guest}, {Gurovich}, {Handberg}, {Hart}, {Hatfield-Dodds}, {Homeier},
  {Hosseinzadeh}, {Jenness}, {Jones}, {Joseph}, {Kalmbach}, {Karamehmetoglu},
  {Ka{l}uszy{'n}ski}, {Kelley}, {Kern}, {Kerzendorf}, {Koch}, {Kulumani},
  {Lee}, {Ly}, {Ma}, {MacBride}, {Maljaars}, {Muna}, {Murphy}, {Norman},
  {O'Steen}, {Oman}, {Pacifici}, {Pascual}, {Pascual-Granado}, {Patil},
  {Perren}, {Pickering}, {Rastogi}, {Roulston}, {Ryan}, {Rykoff}, {Sabater},
  {Sakurikar}, {Salgado}, {Sanghi}, {Saunders}, {Savchenko}, {Schwardt},
  {Seifert-Eckert}, {Shih}, {Jain}, {Shukla}, {Sick}, {Simpson},
  {Singanamalla}, {Singer}, {Singhal}, {Sinha}, {Sip{H{o}}cz}, {Spitler},
  {Stansby}, {Streicher}, {{{S}}umak}, {Swinbank}, {Taranu}, {Tewary},
  {Tremblay}, {Val-Borro}, {Van Kooten}, {Vasovi{'c}}, {Verma}, {de Miranda
  Cardoso}, {Williams}, {Wilson}, {Winkel}, {Wood-Vasey}, {Xue}, {Yoachim},
  {Zhang}, {Zonca}, \& {Astropy Project Contributors}}]{astropy22}
{Astropy Collaboration}, {Price-Whelan}, A.~M., {Lim}, P.~L., {et~al.} 2022,
  apj, 935, 167, \dodoi{10.3847/1538-4357/ac7c74}

\bibitem[{{Bamford} {et~al.}(2009){Bamford}, {Nichol}, {Baldry}, {Land},
  {Lintott}, {Schawinski}, {Slosar}, {Szalay}, {Thomas}, {Torki}, {Andreescu},
  {Edmondson}, {Miller}, {Murray}, {Raddick}, \& {Vandenberg}}]{bamford2009}
{Bamford}, S.~P., {Nichol}, R.~C., {Baldry}, I.~K., {et~al.} 2009, \mnras, 393,
  1324, \dodoi{10.1111/j.1365-2966.2008.14252.x}

\bibitem[{{Benavides} {et~al.}(2020){Benavides}, {Sales}, \&
  {Abadi}}]{benavides2020}
{Benavides}, J.~A., {Sales}, L.~V., \& {Abadi}, M.~G. 2020, \mnras, 498, 3852,
  \dodoi{10.1093/mnras/staa2636}

\bibitem[{{Bertin} \& {Arnouts}(1996)}]{bertin96}
{Bertin}, E., \& {Arnouts}, S. 1996, \aaps, 117, 393

\bibitem[{{Binggeli} {et~al.}(1993){Binggeli}, {Popescu}, \&
  {Tammann}}]{binggeli93}
{Binggeli}, B., {Popescu}, C.~C., \& {Tammann}, G.~A. 1993, \aaps, 98, 275

\bibitem[{{Binggeli} {et~al.}(1985){Binggeli}, {Sandage}, \&
  {Tammann}}]{binggeli85}
{Binggeli}, B., {Sandage}, A., \& {Tammann}, G.~A. 1985, \aj, 90, 1681,
  \dodoi{10.1086/113874}

\bibitem[{{Binggeli} {et~al.}(1987){Binggeli}, {Tammann}, \&
  {Sandage}}]{binggeli87}
{Binggeli}, B., {Tammann}, G.~A., \& {Sandage}, A. 1987, \aj, 94, 251,
  \dodoi{10.1086/114467}

\bibitem[{{Bird} {et~al.}(2010){Bird}, {Harris}, {Blakeslee}, \&
  {Flynn}}]{bird10}
{Bird}, S., {Harris}, W.~E., {Blakeslee}, J.~P., \& {Flynn}, C. 2010, \aap,
  524, A71, \dodoi{10.1051/0004-6361/201014876}

\bibitem[{{Biscardi} {et~al.}(2008){Biscardi}, {Raimondo}, {Cantiello}, \&
  {Brocato}}]{biscardi08}
{Biscardi}, I., {Raimondo}, G., {Cantiello}, M., \& {Brocato}, E. 2008, \apj,
  678, 168, \dodoi{10.1086/587126}

\bibitem[{{Blakeslee}(2012)}]{blake12b}
{Blakeslee}, J.~P. 2012, \apss, 341, 179, \dodoi{10.1007/s10509-012-0997-6}

\bibitem[{{Blakeslee} {et~al.}(2023){Blakeslee}, {Cantiello}, {Hudson},
  {Ferrarese}, {Hazra}, {Jensen}, {Peng}, \& {Raimondo}}]{blakeslee23}
{Blakeslee}, J.~P., {Cantiello}, M., {Hudson}, M.~J., {et~al.} 2023, arXiv
  e-prints, arXiv:2306.15170, \dodoi{10.48550/arXiv.2306.15170}

\bibitem[{{Blakeslee} {et~al.}(2021){Blakeslee}, {Jensen}, {Ma}, {Milne}, \&
  {Greene}}]{blakeslee21}
{Blakeslee}, J.~P., {Jensen}, J.~B., {Ma}, C.-P., {Milne}, P.~A., \& {Greene},
  J.~E. 2021, \apj, 911, 65, \dodoi{10.3847/1538-4357/abe86a}

\bibitem[{{Blakeslee} {et~al.}(2002){Blakeslee}, {Lucey}, {Tonry}, {Hudson},
  {Narayanan}, \& {Barris}}]{blake02}
{Blakeslee}, J.~P., {Lucey}, J.~R., {Tonry}, J.~L., {et~al.} 2002, \mnras, 330,
  443, \dodoi{10.1046/j.1365-8711.2002.05080.x}

\bibitem[{{Blakeslee} {et~al.}(1997){Blakeslee}, {Tonry}, \&
  {Metzger}}]{blake97}
{Blakeslee}, J.~P., {Tonry}, J.~L., \& {Metzger}, M.~R. 1997, \aj, 114, 482,
  \dodoi{10.1086/118488}

\bibitem[{{Blakeslee} {et~al.}(2001){Blakeslee}, {Vazdekis}, \&
  {Ajhar}}]{bva01}
{Blakeslee}, J.~P., {Vazdekis}, A., \& {Ajhar}, E.~A. 2001, \mnras, 320, 193

\bibitem[{{Blakeslee} {et~al.}(2009){Blakeslee}, {Jord{\'a}n}, {Mei},
  {C{\^o}t{\'e}}, {Ferrarese}, {Infante}, {Peng}, {Tonry}, \& {West}}]{blake09}
{Blakeslee}, J.~P., {Jord{\'a}n}, A., {Mei}, S., {et~al.} 2009, \apj, 694, 556,
  \dodoi{10.1088/0004-637X/694/1/556}

\bibitem[{{Blakeslee} {et~al.}(2010){Blakeslee}, {Cantiello}, {Mei},
  {C{\^o}t{\'e}}, {Barber DeGraaff}, {Ferrarese}, {Jord{\'a}n}, {Peng},
  {Tonry}, \& {Worthey}}]{blake10b}
{Blakeslee}, J.~P., {Cantiello}, M., {Mei}, S., {et~al.} 2010, \apj, 724, 657,
  \dodoi{10.1088/0004-637X/724/1/657}

\bibitem[{{Blanton} {et~al.}(2005){Blanton}, {Eisenstein}, {Hogg}, {Schlegel},
  \& {Brinkmann}}]{blanton2005}
{Blanton}, M.~R., {Eisenstein}, D., {Hogg}, D.~W., {Schlegel}, D.~J., \&
  {Brinkmann}, J. 2005, \apj, 629, 143, \dodoi{10.1086/422897}

\bibitem[{{Boselli} \& {Gavazzi}(2006)}]{boselli06}
{Boselli}, A., \& {Gavazzi}, G. 2006, \pasp, 118, 517, \dodoi{10.1086/500691}

\bibitem[{{Boselli} {et~al.}(2014){Boselli}, {Voyer}, {Boissier}, {Cucciati},
  {Consolandi}, {Cortese}, {Fumagalli}, {Gavazzi}, {Heinis}, {Roehlly}, \&
  {Toloba}}]{boselli14}
{Boselli}, A., {Voyer}, E., {Boissier}, S., {et~al.} 2014, \aap, 570, A69,
  \dodoi{10.1051/0004-6361/201424419}

\bibitem[{{Bressan} {et~al.}(2012){Bressan}, {Marigo}, {Girardi}, {Salasnich},
  {Dal Cero}, {Rubele}, \& {Nanni}}]{bressan12}
{Bressan}, A., {Marigo}, P., {Girardi}, L., {et~al.} 2012, \mnras, 427, 127,
  \dodoi{10.1111/j.1365-2966.2012.21948.x}

\bibitem[{{Brocato} {et~al.}(2000){Brocato}, {Castellani}, {Poli}, \&
  {Raimondo}}]{brocato00}
{Brocato}, E., {Castellani}, V., {Poli}, F.~M., \& {Raimondo}, G. 2000, \aaps,
  146, 91

\bibitem[{{Brocato} {et~al.}(1999){Brocato}, {Castellani}, {Raimondo}, \&
  {Romaniello}}]{brocato99}
{Brocato}, E., {Castellani}, V., {Raimondo}, G., \& {Romaniello}, M. 1999,
  \aaps, 136, 65

\bibitem[{{Cantiello} {et~al.}(2011){Cantiello}, {Biscardi}, {Brocato}, \&
  {Raimondo}}]{cantiello11a}
{Cantiello}, M., {Biscardi}, I., {Brocato}, E., \& {Raimondo}, G. 2011, \aap,
  532, A154, \dodoi{10.1051/0004-6361/201116667}

\bibitem[{{Cantiello} {et~al.}(2007{\natexlab{a}}){Cantiello}, {Blakeslee},
  {Raimondo}, {Brocato}, \& {Capaccioli}}]{cantiello07b}
{Cantiello}, M., {Blakeslee}, J., {Raimondo}, G., {Brocato}, E., \&
  {Capaccioli}, M. 2007{\natexlab{a}}, \apj, 668, 130, \dodoi{10.1086/521295}

\bibitem[{{Cantiello} \& {Blakeslee}(2023)}]{cantielloblakesleeH0}
{Cantiello}, M., \& {Blakeslee}, J.~P. 2023, arXiv e-prints, arXiv:2307.03116,
  \dodoi{10.48550/arXiv.2307.03116}

\bibitem[{{Cantiello} {et~al.}(2005){Cantiello}, {Blakeslee}, {Raimondo},
  {Mei}, {Brocato}, \& {Capaccioli}}]{cantiello05}
{Cantiello}, M., {Blakeslee}, J.~P., {Raimondo}, G., {et~al.} 2005, \apj, 634,
  239, \dodoi{10.1086/491694}

\bibitem[{{Cantiello} {et~al.}(2007{\natexlab{b}}){Cantiello}, {Raimondo},
  {Blakeslee}, {Brocato}, \& {Capaccioli}}]{cantiello07a}
{Cantiello}, M., {Raimondo}, G., {Blakeslee}, J.~P., {Brocato}, E., \&
  {Capaccioli}, M. 2007{\natexlab{b}}, \apj, 662, 940, \dodoi{10.1086/517984}

\bibitem[{{Cantiello} {et~al.}(2003){Cantiello}, {Raimondo}, {Brocato}, \&
  {Capaccioli}}]{cantiello03}
{Cantiello}, M., {Raimondo}, G., {Brocato}, E., \& {Capaccioli}, M. 2003, \aj,
  125, 2783, \dodoi{10.1086/375322}

\bibitem[{{Cantiello} {et~al.}(2013){Cantiello}, {Grado}, {Blakeslee},
  {Raimondo}, {Di Rico}, {Limatola}, {Brocato}, {Della Valle}, \&
  {Gilmozzi}}]{cantiello13}
{Cantiello}, M., {Grado}, A., {Blakeslee}, J.~P., {et~al.} 2013, \aap, 552,
  A106, \dodoi{10.1051/0004-6361/201220756}

\bibitem[{{Cantiello} {et~al.}(2018{\natexlab{a}}){Cantiello}, {Blakeslee},
  {Ferrarese}, {C{\^o}t{\'e}}, {Roediger}, {Raimondo}, {Peng}, {Gwyn},
  {Durrell}, \& {Cuillandre}}]{cantiello18ngvs}
{Cantiello}, M., {Blakeslee}, J.~P., {Ferrarese}, L., {et~al.}
  2018{\natexlab{a}}, \apj, 856, 126, \dodoi{10.3847/1538-4357/aab043}

\bibitem[{{Cantiello} {et~al.}(2018{\natexlab{b}}){Cantiello}, {Jensen},
  {Blakeslee}, {Berger}, {Levan}, {Tanvir}, {Raimondo}, {Brocato}, {Alexander},
  {Blanchard}, {Branchesi}, {Cano}, {Chornock}, {Covino}, {Cowperthwaite},
  {D'Avanzo}, {Eftekhari}, {Fong}, {Fruchter}, {Grado}, {Hjorth}, {Holz},
  {Lyman}, {Mandel}, {Margutti}, {Nicholl}, {Villar}, \&
  {Williams}}]{cantiello18gw}
{Cantiello}, M., {Jensen}, J.~B., {Blakeslee}, J.~P., {et~al.}
  2018{\natexlab{b}}, \apjl, 854, L31, \dodoi{10.3847/2041-8213/aaad64}

\bibitem[{{Carlsten} {et~al.}(2019){Carlsten}, {Beaton}, {Greco}, \&
  {Greene}}]{carlsten19}
{Carlsten}, S.~G., {Beaton}, R.~L., {Greco}, J.~P., \& {Greene}, J.~E. 2019,
  \apj, 879, 13, \dodoi{10.3847/1538-4357/ab22c1}

\bibitem[{{Castignani} {et~al.}(2022){Castignani}, {Vulcani}, {Finn}, {Combes},
  {Jablonka}, {Rudnick}, {Zaritsky}, {Whalen}, {Conger}, {De Lucia}, {Desai},
  {Koopmann}, {Moustakas}, {Norman}, \& {Townsend}}]{castignani22}
{Castignani}, G., {Vulcani}, B., {Finn}, R.~A., {et~al.} 2022, \apjs, 259, 43,
  \dodoi{10.3847/1538-4365/ac45f7}

\bibitem[{{Chen} {et~al.}(2013){Chen}, {Kavelaars}, {Gwyn}, {Ferrarese},
  {C{\^o}t{\'e}}, {Jord{\'a}n}, {Suc}, {Cuillandre}, \& {Ip}}]{chen13}
{Chen}, Y.-T., {Kavelaars}, J.~J., {Gwyn}, S., {et~al.} 2013, \apjl, 775, L8,
  \dodoi{10.1088/2041-8205/775/1/L8}

\bibitem[{{Chung} {et~al.}(2020){Chung}, {Yoon}, {Cho}, {Lee}, \&
  {Lee}}]{chung20}
{Chung}, C., {Yoon}, S.-J., {Cho}, H., {Lee}, S.-Y., \& {Lee}, Y.-W. 2020,
  \apjs, 250, 33, \dodoi{10.3847/1538-4365/abb4e6}

\bibitem[{{Cook} {et~al.}(2020){Cook}, {Conroy}, \& {van Dokkum}}]{cook20}
{Cook}, B.~A., {Conroy}, C., \& {van Dokkum}, P. 2020, \apj, 893, 160,
  \dodoi{10.3847/1538-4357/ab83ea}

\bibitem[{{C{\^o}t{\'e}} {et~al.}(2004){C{\^o}t{\'e}}, {Blakeslee},
  {Ferrarese}, {Jord{\'a}n}, {Mei}, {Merritt}, {Milosavljevi{\'c}}, {Peng},
  {Tonry}, \& {West}}]{cote04}
{C{\^o}t{\'e}}, P., {Blakeslee}, J.~P., {Ferrarese}, L., {et~al.} 2004, \apjs,
  153, 223, \dodoi{10.1086/421490}

\bibitem[{{de Vaucouleurs}(1961)}]{deVauc61}
{de Vaucouleurs}, G. 1961, \apjs, 6, 213, \dodoi{10.1086/190064}

\bibitem[{{de Vaucouleurs} \& {de Vaucouluers}(1973)}]{deVauc73}
{de Vaucouleurs}, G., \& {de Vaucouluers}, A. 1973, \aap, 28, 109

\bibitem[{{Dressler}(1980)}]{dressler1980}
{Dressler}, A. 1980, \apj, 236, 351, \dodoi{10.1086/157753}

\bibitem[{{Durrell} {et~al.}(2014){Durrell}, {C{\^o}t{\'e}}, {Peng},
  {Blakeslee}, {Ferrarese}, {Mihos}, {Puzia}, {Lan{\c c}on}, {Liu}, {Zhang},
  {Cuillandre}, {McConnachie}, {Jord{\'a}n}, {Accetta}, {Boissier}, {Boselli},
  {Courteau}, {Duc}, {Emsellem}, {Gwyn}, {Mei}, \& {Taylor}}]{durrell14}
{Durrell}, P.~R., {C{\^o}t{\'e}}, P., {Peng}, E.~W., {et~al.} 2014, \apj, 794,
  103, \dodoi{10.1088/0004-637X/794/2/103}

\bibitem[{{Ellison} {et~al.}(2010){Ellison}, {Patton}, {Simard}, {McConnachie},
  {Baldry}, \& {Mendel}}]{ellison2010}
{Ellison}, S.~L., {Patton}, D.~R., {Simard}, L., {et~al.} 2010, \mnras, 407,
  1514, \dodoi{10.1111/j.1365-2966.2010.17076.x}

\bibitem[{{Ellison} {et~al.}(2009){Ellison}, {Simard}, {Cowan}, {Baldry},
  {Patton}, \& {McConnachie}}]{ellison2009}
{Ellison}, S.~L., {Simard}, L., {Cowan}, N.~B., {et~al.} 2009, \mnras, 396,
  1257, \dodoi{10.1111/j.1365-2966.2009.14817.x}

\bibitem[{{Fantin} {et~al.}(2017){Fantin}, {C{\^o}t{\'e}}, {Hanes}, {Gwyn},
  {Bianchi}, {Ferrarese}, {Cuillandre}, {McConnachie}, \&
  {Starkenburg}}]{fantin17}
{Fantin}, N.~J., {C{\^o}t{\'e}}, P., {Hanes}, D.~A., {et~al.} 2017, \apj, 843,
  53, \dodoi{10.3847/1538-4357/aa7755}

\bibitem[{{Ferrarese} {et~al.}(2012){Ferrarese}, {C{\^o}t{\'e}}, {Cuillandre},
  {Gwyn}, {Peng}, {MacArthur}, {Duc}, {Boselli}, {Mei}, {Erben}, {McConnachie},
  {Durrell}, {Mihos}, {Jord{\'a}n}, {Lan{\c c}on}, {Puzia}, {Emsellem},
  {Balogh}, {Blakeslee}, {van Waerbeke}, {Gavazzi}, {Vollmer}, {Kavelaars},
  {Woods}, {Ball}, {Boissier}, {Courteau}, {Ferriere}, {Gavazzi},
  {Hildebrandt}, {Hudelot}, {Huertas-Company}, {Liu}, {McLaughlin}, {Mellier},
  {Milkeraitis}, {Schade}, {Balkowski}, {Bournaud}, {Carlberg}, {Chapman},
  {Hoekstra}, {Peng}, {Sawicki}, {Simard}, {Taylor}, {Tully}, {van Driel},
  {Wilson}, {Burdullis}, {Mahoney}, \& {Manset}}]{ferrarese12}
{Ferrarese}, L., {C{\^o}t{\'e}}, P., {Cuillandre}, J.-C., {et~al.} 2012, \apjs,
  200, 4, \dodoi{10.1088/0067-0049/200/1/4}

\bibitem[{{Ferrarese} {et~al.}(2016){Ferrarese}, {C{\^o}t{\'e}},
  {S{\'a}nchez-Janssen}, {Roediger}, {McConnachie}, {Durrell}, {MacArthur},
  {Blakeslee}, {Duc}, {Boissier}, {Boselli}, {Courteau}, {Cuillandre},
  {Emsellem}, {Gwyn}, {Guhathakurta}, {Jord{\'a}n}, {Lan{\c{c}}on}, {Liu},
  {Mei}, {Mihos}, {Navarro}, {Peng}, {Puzia}, {Taylor}, {Toloba}, \&
  {Zhang}}]{ferrarese16}
{Ferrarese}, L., {C{\^o}t{\'e}}, P., {S{\'a}nchez-Janssen}, R., {et~al.} 2016,
  \apj, 824, 10, \dodoi{10.3847/0004-637X/824/1/10}

\bibitem[{{Ferrarese} {et~al.}(2020){Ferrarese}, {C{\^o}t{\'e}}, {MacArthur},
  {Durrell}, {Gwyn}, {Duc}, {S{\'a}nchez-Janssen}, {Santos}, {Blakeslee},
  {Boselli}, {Boyer}, {Cantiello}, {Courteau}, {Cuillandre}, {Emsellem},
  {Erben}, {Gavazzi}, {Guhathakurta}, {Huertas-Company}, {Jord{\'a}n},
  {Lan{\c{c}}on}, {Liu}, {Mei}, {Mihos}, {Peng}, {Puzia}, {Roediger}, {Schade},
  {Taylor}, {Toloba}, \& {Zhang}}]{ferrarese20}
{Ferrarese}, L., {C{\^o}t{\'e}}, P., {MacArthur}, L.~A., {et~al.} 2020, \apj,
  890, 128, \dodoi{10.3847/1538-4357/ab339f}

\bibitem[{{Ferrarese et al.}(2024)}]{ferrarese24}
{Ferrarese et al.} 2024, in preparation

\bibitem[{{Fitzpatrick}(1999)}]{fitzpatrick99}
{Fitzpatrick}, E.~L. 1999, \pasp, 111, 63, \dodoi{10.1086/316293}

\bibitem[{{Ftaclas} {et~al.}(1984){Ftaclas}, {Struble}, \&
  {Fanelli}}]{ftaclas84}
{Ftaclas}, C., {Struble}, M.~F., \& {Fanelli}, M.~N. 1984, \apj, 282, 19,
  \dodoi{10.1086/162172}

\bibitem[{{Gavazzi} {et~al.}(1999){Gavazzi}, {Boselli}, {Scodeggio}, {Pierini},
  \& {Belsole}}]{gavazzi1999}
{Gavazzi}, G., {Boselli}, A., {Scodeggio}, M., {Pierini}, D., \& {Belsole}, E.
  1999, \mnras, 304, 595, \dodoi{10.1046/j.1365-8711.1999.02350.x}

\bibitem[{{Greco} {et~al.}(2021){Greco}, {van Dokkum}, {Danieli}, {Carlsten},
  \& {Conroy}}]{greco21}
{Greco}, J.~P., {van Dokkum}, P., {Danieli}, S., {Carlsten}, S.~G., \&
  {Conroy}, C. 2021, \apj, 908, 24, \dodoi{10.3847/1538-4357/abd030}

\bibitem[{Hartigan \& Hartigan(1985)}]{hartigan85}
Hartigan, J.~A., \& Hartigan, P.~M. 1985, The Annals of Statistics, 13, 70 ,
  \dodoi{10.1214/aos/1176346577}

\bibitem[{{Jensen} {et~al.}(2015){Jensen}, {Blakeslee}, {Gibson}, {Lee},
  {Cantiello}, {Raimondo}, {Boyer}, \& {Cho}}]{jensen15}
{Jensen}, J.~B., {Blakeslee}, J.~P., {Gibson}, Z., {et~al.} 2015, \apj, 808,
  91, \dodoi{10.1088/0004-637X/808/1/91}

\bibitem[{{Jensen} {et~al.}(2003){Jensen}, {Tonry}, {Barris}, {Thompson},
  {Liu}, {Rieke}, {Ajhar}, \& {Blakeslee}}]{jensen03}
{Jensen}, J.~B., {Tonry}, J.~L., {Barris}, B.~J., {et~al.} 2003, \apj, 583,
  712, \dodoi{10.1086/345430}

\bibitem[{{Jensen} {et~al.}(2021){Jensen}, {Blakeslee}, {Ma}, {Milne}, {Brown},
  {Cantiello}, {Garnavich}, {Greene}, {Lucey}, {Phan}, {Tully}, \&
  {Wood}}]{jensen21}
{Jensen}, J.~B., {Blakeslee}, J.~P., {Ma}, C.-P., {et~al.} 2021, \apjs, 255,
  21, \dodoi{10.3847/1538-4365/ac01e7}

\bibitem[{{Kim} {et~al.}(2016){Kim}, {Rey}, {Bureau}, {Yoon}, {Chung},
  {Jerjen}, {Lisker}, {Jeong}, {Sung}, {Lee}, {Lee}, \& {Chung}}]{kim16}
{Kim}, S., {Rey}, S.-C., {Bureau}, M., {et~al.} 2016, \apj, 833, 207,
  \dodoi{10.3847/1538-4357/833/2/207}

\bibitem[{{Kim} \& {Lee}(2021)}]{kim21}
{Kim}, Y.~J., \& {Lee}, M.~G. 2021, \apj, 923, 152,
  \dodoi{10.3847/1538-4357/ac2d94}

\bibitem[{{Lee} \& {Jang}(2017)}]{leejang17}
{Lee}, M.~G., \& {Jang}, I.~S. 2017, \apj, 841, 23,
  \dodoi{10.3847/1538-4357/aa6c6a}

\bibitem[{{Lim} {et~al.}(2020){Lim}, {C{\^o}t{\'e}}, {Peng}, {Ferrarese},
  {Roediger}, {Durrell}, {Mihos}, {Wang}, {Gwyn}, {Cuillandre}, {Liu},
  {S{\'a}nchez-Janssen}, {Toloba}, {Sales}, {Guhathakurta}, {Lan{\c{c}}on}, \&
  {Puzia}}]{lim20}
{Lim}, S., {C{\^o}t{\'e}}, P., {Peng}, E.~W., {et~al.} 2020, \apj, 899, 69,
  \dodoi{10.3847/1538-4357/aba433}

\bibitem[{{Lisker} {et~al.}(2018){Lisker}, {Vijayaraghavan}, {Janz},
  {Gallagher}, {Engler}, \& {Urich}}]{lisker2018}
{Lisker}, T., {Vijayaraghavan}, R., {Janz}, J., {et~al.} 2018, \apj, 865, 40,
  \dodoi{10.3847/1538-4357/aadae1}

\bibitem[{{Liu} {et~al.}(2020){Liu}, {C{\^o}t{\'e}}, {Peng}, {Roediger},
  {Zhang}, {Ferrarese}, {S{\'a}nchez-Janssen}, {Guhathakurta}, {Yang}, {Jing},
  {Alamo-Mart{\'\i}nez}, {Blakeslee}, {Boselli}, {Cuilandre}, {Duc}, {Durrell},
  {Gwyn}, {Jord{\'a}n}, {Ko}, {Lan{\c{c}}on}, {Lim}, {Longobardi}, {Mei},
  {Mihos}, {Mu{\~n}oz}, {Powalka}, {Puzia}, {Spengler}, \& {Toloba}}]{liu20}
{Liu}, C., {C{\^o}t{\'e}}, P., {Peng}, E.~W., {et~al.} 2020, \apjs, 250, 17,
  \dodoi{10.3847/1538-4365/abad91}

\bibitem[{{Lokhorst} {et~al.}(2016){Lokhorst}, {Starkenburg}, {McConnachie},
  {Navarro}, {Ferrarese}, {C{\^o}t{\'e}}, {Liu}, {Peng}, {Gwyn}, {Cuillandre},
  \& {Guhathakurta}}]{Lokhorst16}
{Lokhorst}, D., {Starkenburg}, E., {McConnachie}, A.~W., {et~al.} 2016, \apj,
  819, 124, \dodoi{10.3847/0004-637X/819/2/124}

\bibitem[{{Makarov} {et~al.}(2014){Makarov}, {Prugniel}, {Terekhova},
  {Courtois}, \& {Vauglin}}]{makarov14}
{Makarov}, D., {Prugniel}, P., {Terekhova}, N., {Courtois}, H., \& {Vauglin},
  I. 2014, \aap, 570, A13, \dodoi{10.1051/0004-6361/201423496}

\bibitem[{{McLaughlin}(1999)}]{mclaughlin99}
{McLaughlin}, D.~E. 1999, \apjl, 512, L9, \dodoi{10.1086/311860}

\bibitem[{{Mei} {et~al.}(2005{\natexlab{a}}){Mei}, {Blakeslee}, {Tonry},
  {Jord{\' a}n}, {Peng}, {C{\^ o}t{\' e}}, {Ferrarese}, {Merritt},
  {Milosavljevi{\' c}}, \& {West}}]{mei05iv}
{Mei}, S., {Blakeslee}, J.~P., {Tonry}, J.~L., {et~al.} 2005{\natexlab{a}},
  \apjs, 156, 113, \dodoi{10.1086/426544}

\bibitem[{{Mei} {et~al.}(2005{\natexlab{b}}){Mei}, {Blakeslee}, {Tonry},
  {Jord{\' a}n}, {Peng}, {C{\^ o}t{\' e}}, {Ferrarese}, {West}, {Merritt}, \&
  {Milosavljevi{\' c}}}]{mei05v}
---. 2005{\natexlab{b}}, \apj, 625, 121, \dodoi{10.1086/429554}

\bibitem[{{Mei} {et~al.}(2007){Mei}, {Blakeslee}, {C{\^o}t{\'e}}, {Tonry},
  {West}, {Ferrarese}, {Jord{\'a}n}, {Peng}, {Anthony}, \&
  {Merritt}}]{mei07xiii}
{Mei}, S., {Blakeslee}, J.~P., {C{\^o}t{\'e}}, P., {et~al.} 2007, \apj, 655,
  144, \dodoi{10.1086/509598}

\bibitem[{{Mihos} {et~al.}(2022){Mihos}, {Durrell}, {Toloba}, {C{\^o}t{\'e}},
  {Ferrarese}, {Guhathakurta}, {Lim}, {Peng}, \& {Sales}}]{mihos22}
{Mihos}, J.~C., {Durrell}, P.~R., {Toloba}, E., {et~al.} 2022, \apj, 924, 87,
  \dodoi{10.3847/1538-4357/ac35d9}

\bibitem[{{Moresco} {et~al.}(2022){Moresco}, {Amati}, {Amendola}, {Birrer},
  {Blakeslee}, {Cantiello}, {Cimatti}, {Darling}, {Della Valle}, {Fishbach},
  {Grillo}, {Hamaus}, {Holz}, {Izzo}, {Jimenez}, {Lusso}, {Meneghetti},
  {Piedipalumbo}, {Pisani}, {Pourtsidou}, {Pozzetti}, {Quartin}, {Risaliti},
  {Rosati}, \& {Verde}}]{moresco22}
{Moresco}, M., {Amati}, L., {Amendola}, L., {et~al.} 2022, arXiv e-prints,
  arXiv:2201.07241.
\newblock \doarXiv{2201.07241}

\bibitem[{{Muratov} \& {Gnedin}(2010)}]{muratov10}
{Muratov}, A.~L., \& {Gnedin}, O.~Y. 2010, \apj, 718, 1266,
  \dodoi{10.1088/0004-637X/718/2/1266}

\bibitem[{{Neilsen} \& {Tsvetanov}(2000)}]{neilsen20}
{Neilsen}, Eric~H., J., \& {Tsvetanov}, Z.~I. 2000, \apj, 536, 255,
  \dodoi{10.1086/308915}

\bibitem[{{Park} \& {Hwang}(2009)}]{park2009}
{Park}, C., \& {Hwang}, H.~S. 2009, \apj, 699, 1595,
  \dodoi{10.1088/0004-637X/699/2/1595}

\bibitem[{{Paudel} {et~al.}(2013){Paudel}, {Duc}, {C{\^o}t{\'e}}, {Cuillandre},
  {Ferrarese}, {Ferriere}, {Gwyn}, {Mihos}, {Vollmer}, {Balogh}, {Carlberg},
  {Boissier}, {Boselli}, {Durrell}, {Emsellem}, {MacArthur}, {Mei},
  {Michel-Dansac}, \& {van Driel}}]{paudel13}
{Paudel}, S., {Duc}, P.-A., {C{\^o}t{\'e}}, P., {et~al.} 2013, \apj, 767, 133,
  \dodoi{10.1088/0004-637X/767/2/133}

\bibitem[{{Pietrzy{\'n}ski} {et~al.}(2019){Pietrzy{\'n}ski}, {Graczyk},
  {Gallenne}, {Gieren}, {Thompson}, {Pilecki}, {Karczmarek}, {G{\'o}rski},
  {Suchomska}, {Taormina}, {Zgirski}, {Wielg{\'o}rski}, {Ko{\l}aczkowski},
  {Konorski}, {Villanova}, {Nardetto}, {Kervella}, {Bresolin}, {Kudritzki},
  {Storm}, {Smolec}, \& {Narloch}}]{pietrzynski19}
{Pietrzy{\'n}ski}, G., {Graczyk}, D., {Gallenne}, A., {et~al.} 2019, \nat, 567,
  200, \dodoi{10.1038/s41586-019-0999-4}

\bibitem[{{Powalka} {et~al.}(2016){Powalka}, {Lan{\c{c}}on}, {Puzia}, {Peng},
  {Liu}, {Mu{\~n}oz}, {Blakeslee}, {C{\^o}t{\'e}}, {Ferrarese}, {Roediger},
  {S{\'a}nchez-Janssen}, {Zhang}, {Durrell}, {Cuillandre}, {Duc},
  {Guhathakurta}, {Gwyn}, {Hudelot}, {Mei}, \& {Toloba}}]{powalka16}
{Powalka}, M., {Lan{\c{c}}on}, A., {Puzia}, T.~H., {et~al.} 2016, \apjs, 227,
  12, \dodoi{10.3847/0067-0049/227/1/12}

\bibitem[{{Raichoor} {et~al.}(2014){Raichoor}, {Mei}, {Erben}, {Hildebrandt},
  {Huertas-Company}, {Ilbert}, {Licitra}, {Ball}, {Boissier}, {Boselli},
  {Chen}, {C{\^o}t{\'e}}, {Cuillandre}, {Duc}, {Durrell}, {Ferrarese},
  {Guhathakurta}, {Gwyn}, {Kavelaars}, {Lan{\c{c}}on}, {Liu}, {MacArthur},
  {Muller}, {Mu{\~n}oz}, {Peng}, {Puzia}, {Sawicki}, {Toloba}, {Van Waerbeke},
  {Woods}, \& {Zhang}}]{raichoor14}
{Raichoor}, A., {Mei}, S., {Erben}, T., {et~al.} 2014, \apj, 797, 102,
  \dodoi{10.1088/0004-637X/797/2/102}

\bibitem[{{Raimondo}(2009)}]{raimondo09}
{Raimondo}, G. 2009, \apj, 700, 1247, \dodoi{10.1088/0004-637X/700/2/1247}

\bibitem[{{Raimondo} {et~al.}(2005){Raimondo}, {Brocato}, {Cantiello}, \&
  {Capaccioli}}]{raimondo05}
{Raimondo}, G., {Brocato}, E., {Cantiello}, M., \& {Capaccioli}, M. 2005, \aj,
  130, 2625, \dodoi{10.1086/497591}

\bibitem[{{Rodr{\'\i}guez-Beltr{\'a}n}
  {et~al.}(2021){Rodr{\'\i}guez-Beltr{\'a}n}, {Vazdekis}, {Cervi{\~n}o}, \&
  {Beasley}}]{rb21}
{Rodr{\'\i}guez-Beltr{\'a}n}, P., {Vazdekis}, A., {Cervi{\~n}o}, M., \&
  {Beasley}, M.~A. 2021, \mnras, 507, 3005, \dodoi{10.1093/mnras/stab2330}

\bibitem[{{Roediger} {et~al.}(2017){Roediger}, {Ferrarese}, {C{\^o}t{\'e}},
  {MacArthur}, {S{\'a}nchez-Janssen}, {Blakeslee}, {Peng}, {Liu}, {Munoz},
  {Cuillandre}, {Gwyn}, {Mei}, {Boissier}, {Boselli}, {Cantiello}, {Courteau},
  {Duc}, {Lan{\c c}on}, {Mihos}, {Puzia}, {Taylor}, {Durrell}, {Toloba},
  {Guhathakurta}, \& {Zhang}}]{roediger17}
{Roediger}, J.~C., {Ferrarese}, L., {C{\^o}t{\'e}}, P., {et~al.} 2017, \apj,
  836, 120, \dodoi{10.3847/1538-4357/836/1/120}

\bibitem[{{S{\'a}nchez-Janssen} {et~al.}(2016){S{\'a}nchez-Janssen},
  {Ferrarese}, {MacArthur}, {C{\^o}t{\'e}}, {Blakeslee}, {Cuillandre}, {Duc},
  {Durrell}, {Gwyn}, {McConnacchie}, {Boselli}, {Courteau}, {Emsellem}, {Mei},
  {Peng}, {Puzia}, {Roediger}, {Simard}, {Boyer}, \& {Santos}}]{sanchez16}
{S{\'a}nchez-Janssen}, R., {Ferrarese}, L., {MacArthur}, L.~A., {et~al.} 2016,
  \apj, 820, 69, \dodoi{10.3847/0004-637X/820/1/69}

\bibitem[{{S{\'a}nchez-Janssen} {et~al.}(2019){S{\'a}nchez-Janssen},
  {C{\^o}t{\'e}}, {Ferrarese}, {Peng}, {Roediger}, {Blakeslee}, {Emsellem},
  {Puzia}, {Spengler}, {Taylor}, {{\'A}lamo-Mart{\'\i}nez}, {Boselli},
  {Cantiello}, {Cuillandre}, {Duc}, {Durrell}, {Gwyn}, {MacArthur},
  {Lan{\c{c}}on}, {Lim}, {Liu}, {Mei}, {Miller}, {Mu{\~n}oz}, {Mihos},
  {Paudel}, {Powalka}, \& {Toloba}}]{sanchez19}
{S{\'a}nchez-Janssen}, R., {C{\^o}t{\'e}}, P., {Ferrarese}, L., {et~al.} 2019,
  \apj, 878, 18, \dodoi{10.3847/1538-4357/aaf4fd}

\bibitem[{{Schlegel} {et~al.}(1998){Schlegel}, {Finkbeiner}, \&
  {Davis}}]{sfd98}
{Schlegel}, D.~J., {Finkbeiner}, D.~P., \& {Davis}, M. 1998, \apj, 500, 525,
  \dodoi{10.1086/305772}

\bibitem[{{Spriggs} {et~al.}(2021){Spriggs}, {Sarzi}, {Gal{\'a}n-de Anta},
  {Napiwotzki}, {Viaene}, {Nedelchev}, {Coccato}, {Corsini}, {Fahrion},
  {Falc{\'o}n-Barroso}, {Gadotti}, {Iodice}, {Lyubenova},
  {Mart{\'\i}n-Navarro}, {McDermid}, {Morelli}, {Pinna}, {van de Ven}, {de
  Zeeuw}, \& {Zhu}}]{spriggs21}
{Spriggs}, T.~W., {Sarzi}, M., {Gal{\'a}n-de Anta}, P.~M., {et~al.} 2021, \aap,
  653, A167, \dodoi{10.1051/0004-6361/202141314}

\bibitem[{{Taylor}(2005)}]{taylor05}
{Taylor}, M.~B. 2005, in Astronomical Society of the Pacific Conference Series,
  Vol. 347, Astronomical Data Analysis Software and Systems XIV, ed.
  P.~{Shopbell}, M.~{Britton}, \& R.~{Ebert}, 29

\bibitem[{{Tonry} \& {Schneider}(1988)}]{ts88}
{Tonry}, J., \& {Schneider}, D.~P. 1988, \aj, 96, 807, \dodoi{10.1086/114847}

\bibitem[{{Tonry}(1991)}]{tonry91}
{Tonry}, J.~L. 1991, \apjl, 373, L1, \dodoi{10.1086/186037}

\bibitem[{{Tonry} {et~al.}(1990){Tonry}, {Ajhar}, \& {Luppino}}]{tal90}
{Tonry}, J.~L., {Ajhar}, E.~A., \& {Luppino}, G.~A. 1990, \aj, 100, 1416,
  \dodoi{10.1086/115606}

\bibitem[{{Tonry} {et~al.}(1997){Tonry}, {Blakeslee}, {Ajhar}, \&
  {Dressler}}]{tonry97}
{Tonry}, J.~L., {Blakeslee}, J.~P., {Ajhar}, E.~A., \& {Dressler}, A. 1997,
  \apj, 475, 399, \dodoi{10.1086/303576}

\bibitem[{{Tonry} {et~al.}(2000){Tonry}, {Blakeslee}, {Ajhar}, \&
  {Dressler}}]{tonry00}
---. 2000, \apj, 530, 625, \dodoi{10.1086/308409}

\bibitem[{{Tonry} {et~al.}(2001){Tonry}, {Dressler}, {Blakeslee}, {Ajhar},
  {Fletcher}, {Luppino}, {Metzger}, \& {Moore}}]{tonry01}
{Tonry}, J.~L., {Dressler}, A., {Blakeslee}, J.~P., {et~al.} 2001, \apj, 546,
  681, \dodoi{10.1086/318301}

\bibitem[{{Tully}(1982)}]{tully82}
{Tully}, R.~B. 1982, \apj, 257, 389, \dodoi{10.1086/159999}

\bibitem[{{Wang} {et~al.}(2018){Wang}, {Pearce}, {Knebe}, {Yepes}, {Cui},
  {Power}, {Arth}, {Gottl{\"o}ber}, {De Petris}, {Brown}, \& {Feng}}]{wang18}
{Wang}, Y., {Pearce}, F., {Knebe}, A., {et~al.} 2018, \apj, 868, 130,
  \dodoi{10.3847/1538-4357/aae52e}

\bibitem[{{West} \& {Blakeslee}(2000)}]{west00}
{West}, M.~J., \& {Blakeslee}, J.~P. 2000, \apjl, 543, L27,
  \dodoi{10.1086/318177}

\bibitem[{{Whitmore} {et~al.}(1995){Whitmore}, {Sparks}, {Lucas}, {Macchetto},
  \& {Biretta}}]{whitmore95b}
{Whitmore}, B.~C., {Sparks}, W.~B., {Lucas}, R.~A., {Macchetto}, F.~D., \&
  {Biretta}, J.~A. 1995, \apjl, 454, L73, \dodoi{10.1086/309788}

\bibitem[{{Worthey}(1993)}]{worthey93a}
{Worthey}, G. 1993, \apj, 409, 530, \dodoi{10.1086/172684}

\end{thebibliography}
\bibliographystyle{aasjournal}
\bibstyle{aasjournal}



\end{document}